\documentclass[times,tighten]{aastex631} % , twocolappendix, twocolumn

\usepackage{mathrsfs}
\usepackage{amsmath}

\newcommand{\alphaeff}{\alpha_{\rm eff}}
\newcommand{\alphaAMS}{\alpha_{\rm AMS}}
\newcommand{\cs}{c_{\rm s}}
\newcommand{\dotMp}{\dot{\mathscr{M}}_{\rm p}}
\newcommand{\dotMs}{\dot{\mathscr{M}}_{\rm s}}
\newcommand{\dotMgrow}{\dot{\mathscr{M}}_{\rm grow}}
\newcommand{\dotJ}{\dot{\mathcal{J}}}
\newcommand{\Eout}{E_{\rm out}}
\newcommand{\facc}{f_{\rm acc}}
\newcommand{\Lout}{L_{\rm out}}
\newcommand{\MTot}{M_{\rm Tot}}
\newcommand{\Mout}{M_{\rm out}}
\newcommand{\Mp}{M_{\rm p}}
\newcommand{\Ms}{M_{\rm s}}
\newcommand{\Msun}{M_\odot}
\newcommand{\OmgK}{\Omega_{\rm K}}
\newcommand{\rg}{r_{\rm g}}
\newcommand{\rmd}{{\rm d}}
\newcommand{\rout}{r_{\rm out}}
\newcommand{\rtr}{r_{\rm tr}}
\newcommand{\RBon}{R_{\rm Bon}}
\newcommand{\Rg}{R_{\rm g}}
\newcommand{\Rmax}{R_{\rm max}}
\newcommand{\Rmin}{R_{\rm min}}
\newcommand{\RHill}{R_{\rm Hill}}
\newcommand{\Rout}{R_{\rm out}}
\newcommand{\sigT}{\sigma_{\rm T}}
\newcommand{\Tc}{T_{\rm c}}
\newcommand{\tout}{t_{\rm out}}
\newcommand{\trej}{t_{\rm rej}}
\newcommand{\tmig}{t_{\rm mig}}
\newcommand{\tvis}{t_{\rm vis}}
\newcommand{\vout}{v_{\rm out}}
\newcommand{\Xout}{X_{\rm out}}

\def\ihep{State Key Laboratory of Particle Astrophysics, Institute of High Energy Physics,
Chinese Academy of Sciences, 19B Yuquan Road, Beijing 100049, China}

\def\AstroUCAS{School of Astronomy and Space Sciences, University of Chinese Academy of Sciences, 
19A Yuquan Road, Beijing 100049, China}

\def\naoc{National Astronomical Observatory of China, 20A Datun Road, Beijing 100020, China}

\shorttitle{Accretion-modified Stars in AGNs}
\shortauthors{Liu et al.}

\begin{document}
\title{\large\bf
Accretion-modified Stars in Accretion Disks of Active Galactic Nuclei: Contribution to AGN disk viscosity}

\author[0000-0003-3086-7804]{Jun-Rong Liu}
\affil{\ihep}
\email{liujunrong@ihep.ac.cn}
% \affil{\PhyUCAS}

\author[0000-0001-9449-9268]{Jian-Min Wang}
\affil{\ihep}
\affil{\AstroUCAS}
\affil{\naoc}
\email{wangjm@ihep.ac.cn}

\author[0000-0001-7584-6236]{Hua Feng}
\affil{\ihep}
% \email{hfeng@ihep.ac.cn}

% \linenumbers

\begin{abstract}
It is widely believed that stellar-mass black holes (sMBHs) exist within the accretion disks of active galactic nuclei (AGN), forming a distinct population termed ``accretion-modified star" (AMS).
Gas from the dense disk accretes onto these AMSs, dissipating substantial gravitational energy through a mini-disk around the sMBHs,
which drives powerful outflows that interact with the surrounding disk gas.
In this study, we investigate two scenarios for AMS accretion: episodic Bondi explosions with hyper-Eddington accretion (Scenario A) and steady Eddington accretion (Scenario B).
These outflows generate turbulence, facilitating outward angular momentum transport in the AGN disk via shock interactions and angular momentum exchange.
We explore a broad parameter space-spanning the central supermassive black hole (SMBH) mass ($M_{\rm p}$), dimensionless accretion rate ($\dot{\mathscr{M}}_{\rm p}$), sMBH mass function, and spatial distribution-to calculate the effective viscosity parameter $\alpha_{\rm AMS}$.
Our analysis reveals the scaling relations $\alpha_{\rm AMS}\propto\zeta M_{\rm p}^2\dot{\mathscr{M}}_{\rm p}$ for Scenario A and $\alpha_{\rm AMS}\propto\zeta M_{\rm p}^{1.5}\dot{\mathscr{M}}_{\rm p}^{0.1}$ for Scenario B,
where $\zeta$ denotes the ratio of total sMBH mass to the SMBH disk mass.
For $\zeta={0.01}$ and $M_{\rm p}=10^8 M_\odot$, $\alpha_{\rm AMS}$ ranges from $\sim{3\times10^{-4}}$ to ${0.01}$ (Scenario A) and $\sim{3\times10^{-3}}$ to ${0.04}$ (Scenario B) from the inner to outer disk regions.
These results demonstrate that AMS feedback provides an efficient mechanism for angular momentum transport in AGN disks.  
\end{abstract}

\section{Introduction}

It is well established that active galactic nuclei (AGNs) are powered by gas accretion onto center supermassive black holes \citep[SMBHs, e.g.,][]{Salpeter1964, Zeldovich1964}.
In the accretion disk scenario \citep[e.g.,][]{Lynden-Bell1969, Shakura1973}, gas gradually moves inward as angular momentum is transported outward through viscous stresses \citep[][]{Lynden-Bell1974}.
This viscosity plays a crucial role in converting gravitational energy into radiation by heating the disk gas.
The origin of viscosity in AGN disks is widely attributed to turbulence,
which may be driven by a variety of mechanisms, including the magnetorotational instability \citep[MRI; ][]{Balbus1991,Mishra2020,Begelman2024}, gravitational instability \citep[GI; ][]{Goodman2003, Chen2023YiXian} in self-gravitating regions, the Rayleigh-Taylor instability \citep{Kaisig1992,Marshall2018}, or supernova explosions \citep[][]{Rozyczka1995,Wang2009,Wang2010,Yan2010,Moranchel2021}.

Motivated by the observation of high metallicity in broad-line regions \citep[BLRs;][]{Hamann1999} and the detection of gravitational waves (GWs) from binary black holes (BBHs) mergers involving black holes heavier than typical stellar-mass black hole \citep[sMBH; e.g.,][]{Abbott2020},
it has been recognized that stellar objects can exist in AGN disks \citep[][]{Artymowicz1993,Cheng1999,Graham2020,Graham2023} {and bulges \citep{Zhai2025}}.
The evolution of stars in AGN disks leads primarily to two outcomes, an ``immortal" state or runaway accretion, depending on their location in the disk and the local accretion rate \citep{Ali2023,Dittmann2025, Fabj2025,Xu2025}.
The latter pathway produces energetic transients such as supernovae and $\gamma$-ray bursts (GRBs), leaving behind neutron stars or black holes \citep{Li2023,She2025}.
In addition to in situ evolution, black holes can also be captured from the nuclear stellar cluster and enter the disk \citep{Rowan2025, Whitehead2025b}.
The dense AGN disk environment provides ideal conditions for these seed black holes to grow into the massive black holes observed in exceptional GW events, such as famous GW190521 \citep{Abbott2020,Graham2020}, GW190814 \citep{Abbott2020b,Yang2025}, and the recent GW231123 \citep{LIGO2025,Delfavero2025}.
When two sMBHs come sufficiently close to overcome the tidal force of the central SMBH, they form a BBH \citep{Wang2021b,Whitehead2025}.
These BBHs subsequently interact with the surrounding gas \citep{Dittmann2025b,Fabj2025b,Kishor2025}, the central SMBH \citep{Postiglione2025}, and potentially a third sMBH \citep{Fabj2024,Wang2025a,Wang2025b,Samsing2025}.

Many efforts {have} been made to theoretically study the properties of embedded sMBHs in AGN disks and identify their electromagnetic observational signatures.
When these sMBHs orbit the central SMBH, they form extreme mass-ratio inspiral (EMRI) systems \citep{Abhishek2025,Cheng2025,Duque2025,Sun2025}.
Applied to Sgr A$^*$, such systems can manifest as hot spots and have successfully reproduced the observed infrared light curves and spectral energy distribution \citep[SED,][]{Wang2023SgrA}.
In the $\gamma$-ray band, sMBHs are expected to produce GRB via relativistic jets \citep{Zhang2024,Kang2025,Yuan2025}.
Several candidate events have been reported, such as GW150914 associated with a bright $\gamma$-ray event \citep{Tagawa2023BBH}, GRB 191019A \citep{Lazzati2023,Levan2023}, and GW S241125n associated with a $\gamma$-ray counterpart \citep{Zhang2025}.
In optical and UV bands, the interaction between relativistic jets and the disk gas can form a jet-cocoon structure that emits thermal radiation \citep{Chen2025},
offering a possible explanation for observed AGN optical flares \citep{He2025}.
Additionally, magnetic reconnection driven by sMBHs may account for short-term AGN variability \citep{Xing2025}.
In radio observations, the AGN disk environment has been invoked to explain the large dispersion measure and rotation measure of bright fast radio bursts \citep{Zhao2024}.

The sMBHs embedded in AGN disks often experience hyper-Eddington accretion, significantly exceeding typical super-Eddington accretion rates \citep[e.g.,][]{Wang2021a, Wang2021b, Liu2024, Luo2025,Jiao2025} within the framework of Bondi accretion \citep{Bondi1952}.
This intense accretion plays a critical role in sMBH evolution, leading such objects to be referred to as ``accretion-modified star" \citep[AMS,][]{Wang2021a}.
AMSs are expected to launch powerful outflows \citep[e.g.,][]{Wang2021a, Tagawa2022, Chen2023, Liu2024} or Blandford-Znajek jets \citep[e.g.,][]{Wang2021b, Tagawa2023} as mechanical feedback.
These outflows or jets collide with the AGN disk, producing shocks that heat the surrounding gas and accelerate electrons to relativistic speeds \citep{Blandford1987}.
The process gives rise to additional thermal and non-thermal emission \citep[][]{Wang2021a, Tagawa2023, Liu2024},
which superimpose on the intrinsic AGN radiation.
For instance, collective thermal emission from AMS populations can influence the SED of the SMBH disk and may help explain the shallow spectral shape seen in quasar composite spectra \citep[][]{Zhou2024}.
Meanwhile, $\gamma$-rays originating from AMSs could be detectable in radio-quiet AGNs \citep{Liu2025a,Liu2025b},
where jet-related emission is relatively subdued.

{This work represents the fifth in a series of papers dedicated to exploring the behavior of black holes embedded in AGN disks \citep{Wang2021a,Wang2021b,Wang2023SgrA,Liu2024}.}
In this study, we propose a mechanism whereby outflows from AMSs facilitate outward angular momentum transport in AGN disks through momentum exchange between shocks and disk gas, a process analogous to that invoked in supernova feedback \citep{Rozyczka1995,Moranchel2021}.
This mechanism contributes significantly to the viscosity of the AGN disk,
{particularly in the outer regions, and thus offers a potential solution to the long-standing low-viscosity problem.
Moreover, sufficiently abundant AMS populations can heat the AGN accretion disk, alter its density and temperature profiles \citep{Wang2025d,Epstein2025}, and help maintain gravitational stability in the outer regions of a Shakura-Sunyaev disk.}

The paper is organized as follows:
In \S\,\ref{sect:model}, we describe in detail the AMS models and their feedback, including episodic accretion scenario associated with Bondi explosion (\S\,\ref{sect:Bondi}) and steady accretion scenario (\S\,\ref{sect:steady}).
In \S\,\ref{sect:AMtransfer}, we calculate the angular momentum transfer rate of the AGN disk driven by strong outflows and present the resulting effective viscosity parameter $\alphaAMS$ across a typical parameter space, taking into account different sMBH mass functions and spatial distributions.
In \S\,\ref{sect:discussion}, we summarize the main results and provide a general discussion.
In Appendix~\ref{appdix}, we present the radial profiles of the viscosity parameter as a function of SMBH mass and accretion rate.

\section{Model}
\label{sect:model}

In the context of accretion in the SMBH disk, two distinct physical scenarios are commonly discussed.
The first involves the classical Eddington limit, in which isotropic radiation pressure balances gravitational forces, thus inhibiting further accretion onto the central black hole.
The situation changes considerably, however, in the presence of anisotropic accretion, such as when an accretion disk forms around the black hole \citep[e.g.,][]{Kato2008}.
In such a configuration, matter can accrete at rates significantly exceeding the Eddington limit, enabling not only super-Eddington but even hyper-Eddington accretion flows \citep[e.g.,][]{Abramowicz1980,Wang2021a}.
In this paper, we examine both of these accretion regimes in the context of AMSs and evaluate their role in driving viscosity within AGN disk.

We adopt the standard disk model of \citet{Shakura1973} and investigate the viscosity contributed by outflows driven from AMSs.
We consider a sMBH of mass $\Ms$ embedded in the AGN disk and orbiting a {central} SMBH of mass $\Mp$ with an accretion rate $\dot \Mp$.
Here, the subscript ``p" and ``s" denote ``primary" and ``secondary", respectively.
The dimensionless accretion rate of the SMBH is defined as $\dotMp \equiv \dot \Mp c^2/L_{\rm Edd,\,p}$,
where $L_{\rm Edd,\,p}=4\pi G\Mp m_{\rm p}c/\sigT$ is the Eddington luminosity of the SMBH,
$G$ is the gravitational constant,
$m_{\rm p}$ is the proton mass,
$c$ is the speed of light,
and $\sigT$ is the Thomson scattering cross section.
We explore the parameter space with $\Mp=10^7, 10^8, 10^9\Msun$ and $\dotMp=0.1,1,10$.
The viscosity parameter of the AGN disk is taken as $\alpha=0.1$.

\begin{figure*}
\centering
\includegraphics[width=\linewidth, trim=10 0 0 0]{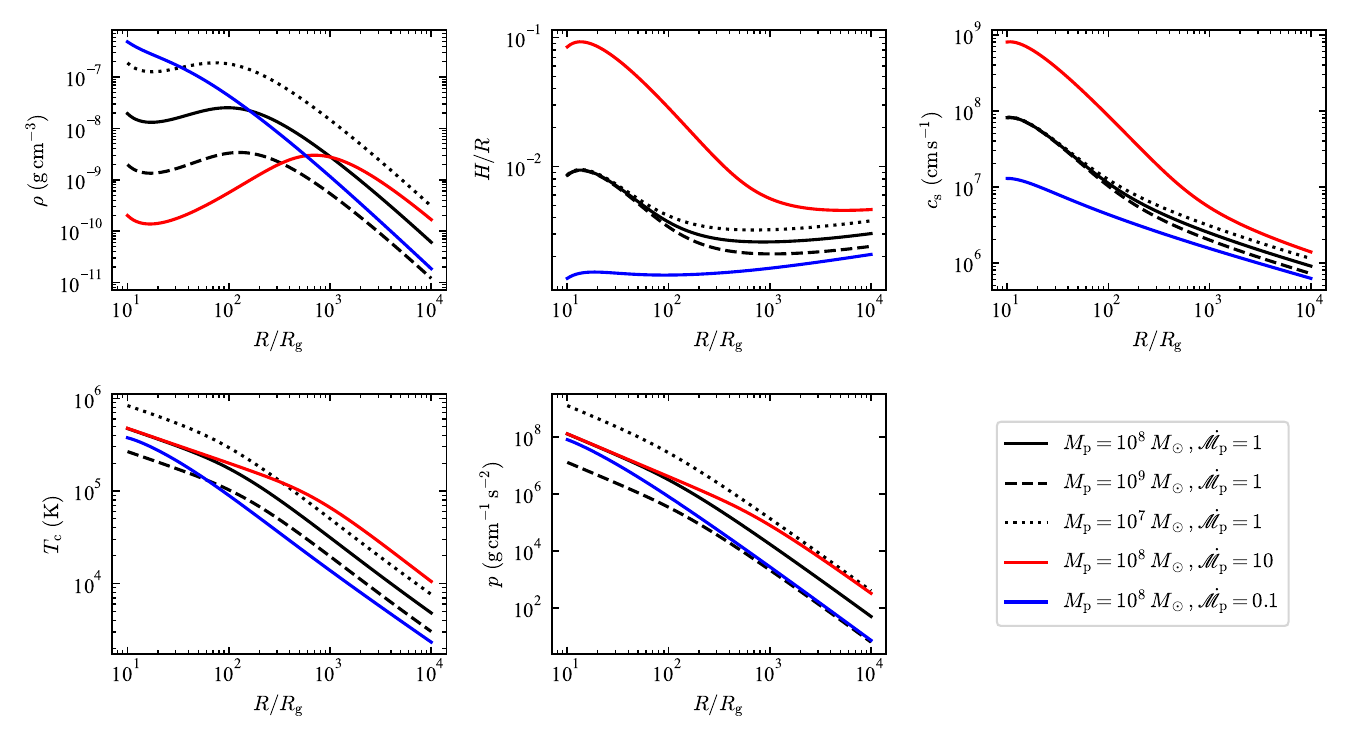}
\caption{Radial distribution of the gas density $\rho$, half-thickness $H$,  sound speed $\cs$, {midplane} temperature $\Tc$, and total pressure $p$ for a standard disk model, shown for different SMBH masses ($\Mp$) and accretion rates ($\dotMp$).}
\label{fig:disk}
\end{figure*}

Figure~\ref{fig:disk} presents the radial profiles of density $\rho$, half-thickness $H$, sound speed $\cs$, midplane temperature $\Tc$, and total pressure $p$ for the standard accretion disk model \citep{Shakura1973,Kato2008}.
It should be noted, however, that the SED of the disk could be modified when radiation from AMS is taken into account \citep{Liu2024, Zhou2024}. 
Furthermore, the outer disk structure may be significantly influenced if the energy released by AMS becomes comparable to the local viscous energy dissipation \citep{Epstein2025,Wang2025c}, in a manner analogous to the feedback-dominated accretion flows discussed in \cite{Gilbaum2022}.
In the present study, we adopt the standard Shakura-Sunyaev disk model and neglect the feedback from AMS energy injection on the disk's temperature and density structure.
This simplification is justifiable when the total sMBH mass is substantially smaller than that of the central SMBH (see mass function of sMBHs in \S\,\ref{sect:NUM})
{and the viscosity contributed by AMS remains below 0.1 (see result in \S\,\ref{sect:result})}.
However, in cases where the total sMBH mass approaches or exceeds the SMBH mass, a self-consistent ``AMS-AGN disk” model would be required to reassess the structural parameters illustrated in Figure~\ref{fig:disk}.
The development of such a model lies beyond the scope of this paper and remains a subject for future investigation.

\subsection{Scenario A: Episodic accretion for Bondi explosion}
\label{sect:Bondi}

\begin{figure*}
\centering
\includegraphics[width= 0.8\textwidth,trim= 0 100 0 0]{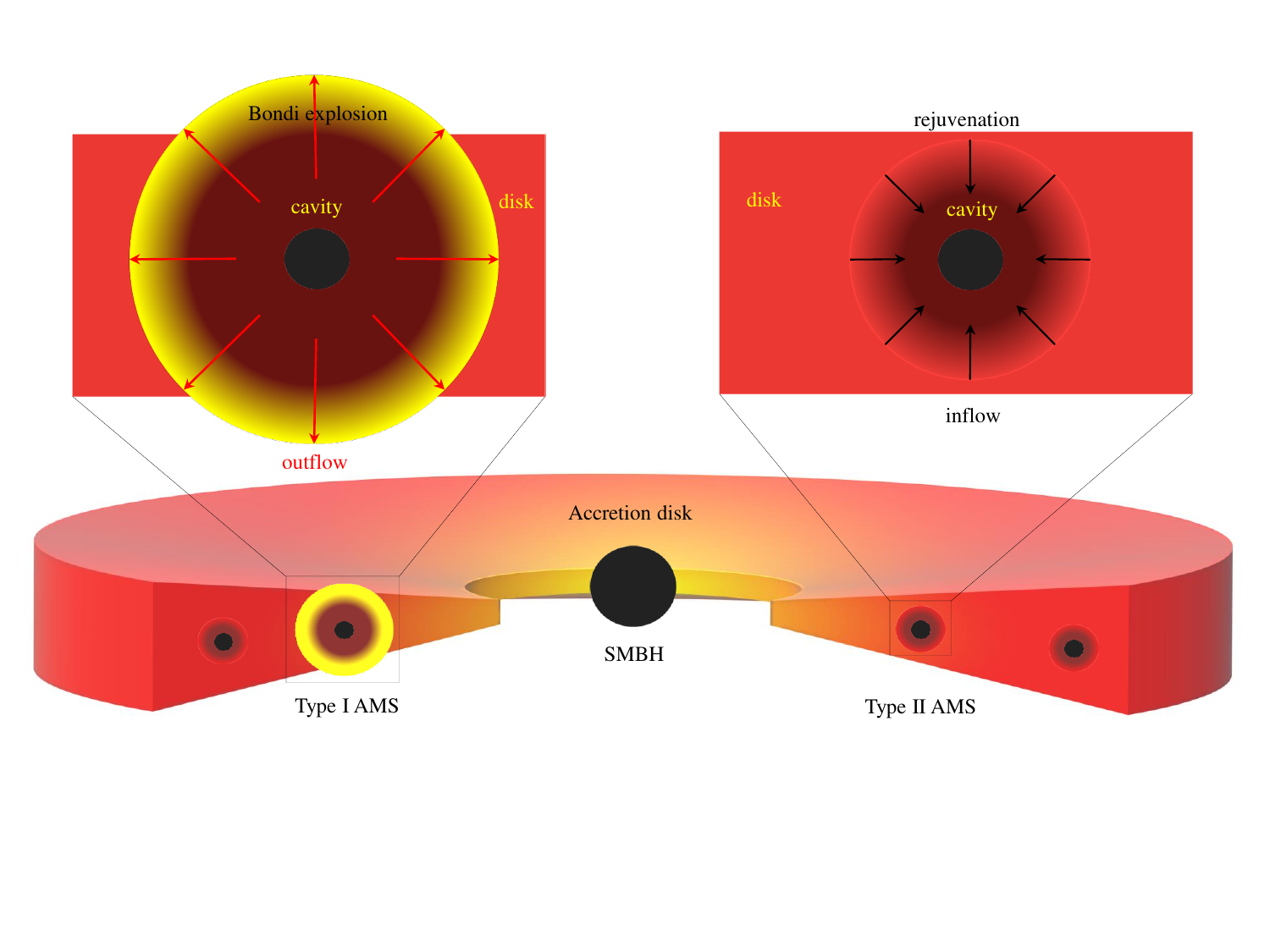}
\caption{Schematic illustration of AMSs in an AGN accretion disk.
The upper left panel depicts a Type I AMS during the Bondi explosion phase,
in which a strong outflow interacts with the surrounding disk gas.
The upper right panel shows a Type II AMS in the subsequent rejuvenation phase, characterized by a low-density cavity.
The bottom panel illustrates the general scenario of sMBHs embedded in the dense AGN disk, accreting gas and forming AMSs.
Gas density is color-coded: yellow (high), red (medium), and brown (low).}
\label{fig:cartoon}
\end{figure*}

Figure~\ref{fig:cartoon} illustrates two types of AMS, Type I (Bondi explosion phase) and Type II (rejuvenation phase), orbiting the central SMBH in Keplerian motion within the AGN accretion disk \citep{Wang2021b}.
Type I AMSs undergo hyper-Eddington accretion fueled by the dense gas of the AGN disk \citep[e.g.,][]{Wang2021a, Chen2023, Liu2024},
producing strong outflows as a mechanical feedback \citep[e.g.,][]{Cao2022, Liu2024}.
When the outflow velocity surpasses the local sound speed, it drives shocks into the surrounding gas \citep[e.g.,][]{Wang2021a, Liu2024}.
These shocks heat the gas and create a low-density cavity (brown region in the figure).
The shocked gas radiates thermal blackbody radiation,
while electrons accelerated to relativistic speeds by the shocks produce non-thermal emission via synchrotron radiation and inverse Compton scattering \citep{Liu2024}.
This violent energy-release process, powered by Bondi accretion, was originally termed a ``Bondi explosion" in \cite{Wang2021a} and has been adopted in their series papers.
After a Type I AMS forms a cavity, the gas within it cools.
Low and medium density gas from the surroundings subsequently refills the cavity under pressure, marking the rejuvenation phase (Type II AMS).
During this stage, the accretion rate drops sharply and the AMS enters a low-activity state.
The rejuvenation time is generally much longer than the outflow expansion timescale,
so Type II AMSs dominate the lifetime of an AMS.
This prevents runaway growth of sMBHs, thereby preserving the AGN disk structure.
In this work, we focus primarily on the Bondi explosion process of Type I AMSs due to their significantly greater energy release compared to Type II AMSs.

We consider a sMBH corotating with the AGN disk, as shown in Figure~\ref{fig:cartoon}.
Its accretion rate is given by the Hoyle-Lyttleton-Bondi formulation \citep{Hoyle1939, Bondi1952}, modified in thin disks \citep{Kocsis2011},
\begin{equation}
\label{eq:dotMs}
\dot{M}_{\rm s}
= \displaystyle \frac{4\pi G^{2}\Ms^{2}\rho}{\left({\cs^2+\Delta v^2}\right)^{3/2}} f_{\rm Hill} f_{\rm H},
\end{equation}
where $f_{\rm Hill}=\min\left\{\RHill/\RBon,\,1\right\}$ is the reduction factor due to tidal effects,
$f_{\rm H}=\min\left\{H/\RBon,\,1\right\}$ accounts for the limited $H$ of the SMBH disk,
and $\RHill=(\Ms/3\Mp)^{1/3}R$ is the Hill radius of the sMBH.
The Bondi radius is defined as \citep[][]{Edgar2004}
\begin{equation}\label{eq:RBon}
\RBon=\frac{G\Ms}{\cs^2}.
\end{equation}
We introduce the dimensionless accretion rate of the AMS as
$\dotMs \equiv \dot{M}_{\rm s}c^2/L_{\rm Edd,\,s}$,
where $L_{\rm Edd,\,s}= 4\pi G\Ms m_{\rm p}c/\sigT$ is the Eddington luminosity of the sMBH.
The relative velocity between the sMBH and the surrounding gas is taken as $\Delta v=\OmgK\times\min\{\RBon,\,\RHill\}/2$,
with $\OmgK=\sqrt{G\Mp/R^3}$ being the Keplerian angular velocity.
While the characteristic accretion scale in isolation is typically the Bondi radius $\RBon$ \citep[e.g.,][]{Edgar2004},
within an AGN disk the effective accretion region is often limited by the smaller Hill radius or disk height \citep{Kocsis2011, Liu2024}.

\begin{figure*}
\centering
\includegraphics[width=\linewidth, trim=10 0 0 0]{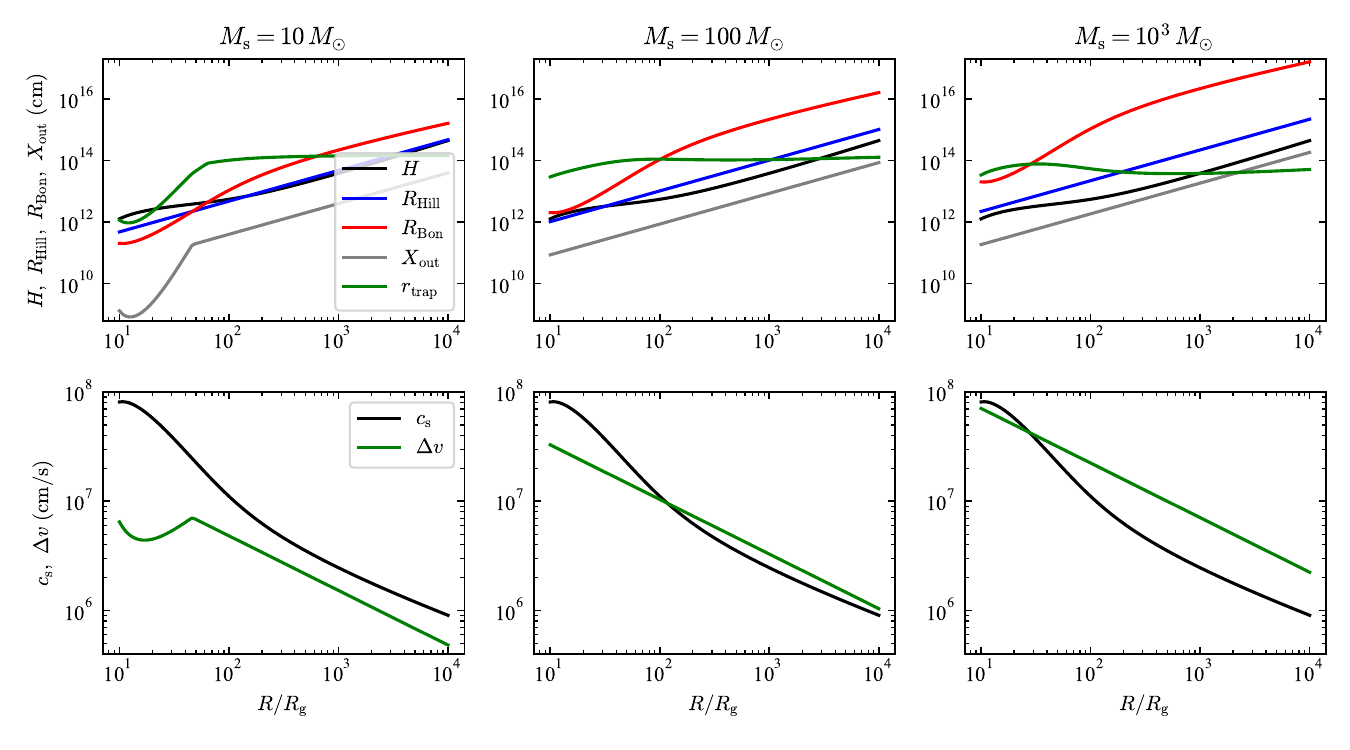}
\caption{Top: The disk height $H$, the Hill radius $\RHill$, the Bondi radius $\RBon$, and the size of the sMBH disk $\Xout$ as functions of the disk radius $R$.
Bottom: The sound speed $\cs$ and the differential velocity $\Delta v$ between the sMBH and the disk gas.
The left, middle, and right panels correspond to sMBH masses of 10$\Msun$, 100$\Msun$, and 1000 $\Msun$, respectively.
The SMBH mass and accretion rate are fixed at $\Mp=10^8\Msun$ and $\dotMp=1$.}
\label{fig:HR}
\end{figure*}

As shown in the upper three panels of Figure~\ref{fig:HR}, $H<\RBon$ and $\RHill<\RBon$ hold in most of the parameter space, except at small disk radii ($R/\Rg < 100$) and for low sMBH masses ($10 \Msun$),
where $\Rg=G\Mp/c^2$ is the gravitational radius and $\Msun$ denotes the solar mass.
These inequalities indicate that the accretion region of an sMBH in an AGN disk is primarily constrained by two factors, i.e., geometry effect ($H$) and tidal effect ($\RHill$).
Even in the outer disk, where $H$ is larger, it remains smaller than $\RBon$, continuing to restrict the accretion regions.
Although increasing $\Ms$ or $R$ {enlarges} $\RHill$, it still does not exceed $\RBon$ in most cases.
Only when both $R$ and $\Ms$ are smaller, where $\RHill > \RBon$ and $H > \RBon$, can both geometry effect and tidal limitations be neglected.
On the other hand, for more massive sMBHs in the outer disk, the accretion increasingly deviates from spherical symmetry.
In addition, the differential velocity between the sMBH and the disk gas suppresses the accretion rate.
The lower three panels of Figure~\ref{fig:HR} compare the relevant velocities and illustrate the resulting suppression of accretion due to this relative motion.
For low-mass sMBHs ($10\,\Msun$), $\cs > \Delta v$, implying subsonic relative motion and favoring spherical Bondi accretion.
In contrast, for higher sMBH masses ($R/\Rg \gtrsim 100$) in the outer disk, $\Delta v$ becomes comparable to or greater than $\cs$, particularly for $\Ms = 10^3 \Msun$.
In such cases, the supersonic relative motion disrupts spherical inflow and significantly reduces $\dot{M}_{\rm s}$.

\begin{figure*}
\centering
\includegraphics[width=\linewidth]{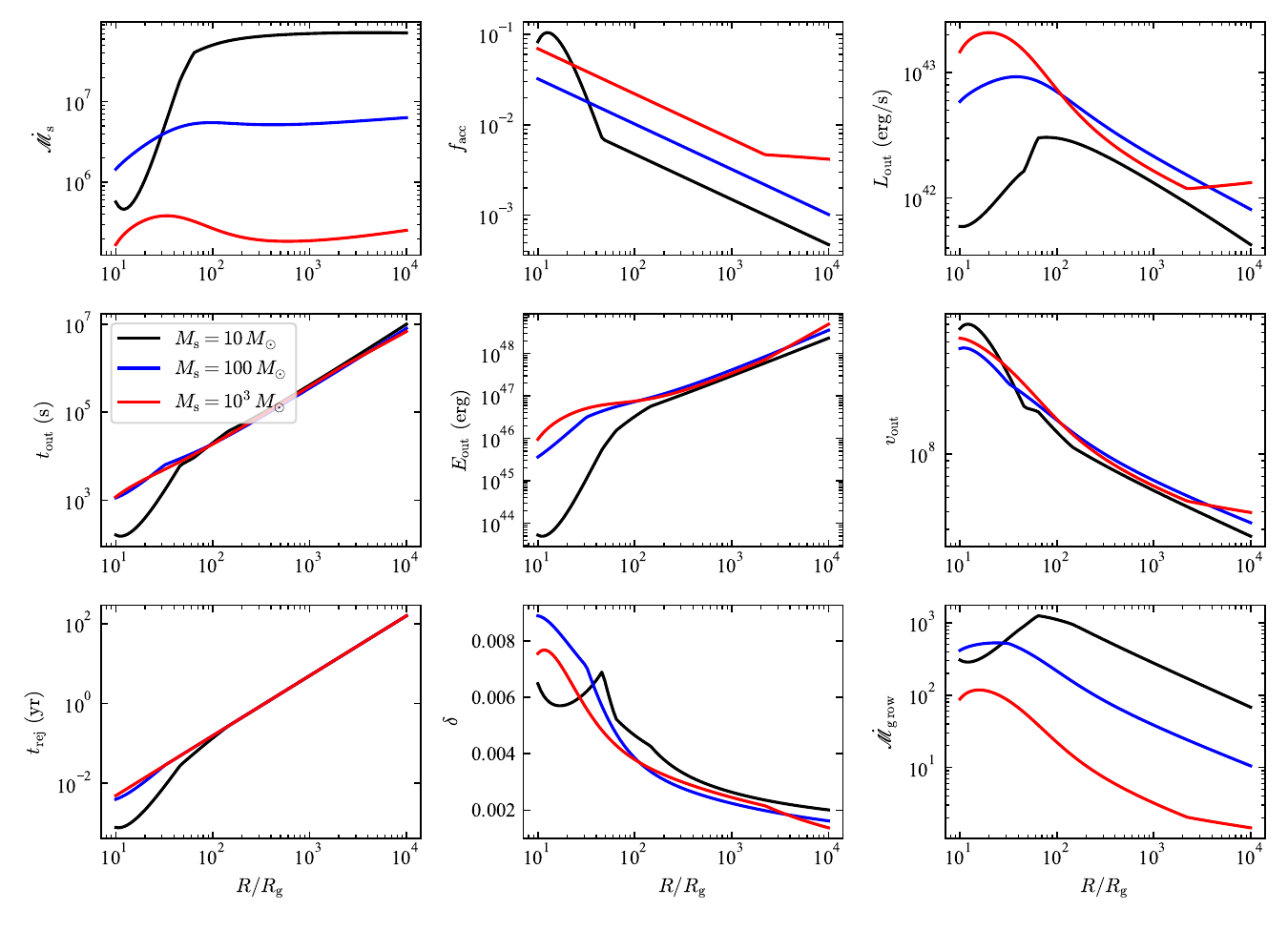}
\caption{Radial profiles of key quantities in the Bondi explosion scenario (Scenario A),
showing their variation with disk radius and sMBH mass ($\Ms$).
Quantities include 
$\dotMs$ (dimensionless sMBH accretion rate),
$\facc$ (accretion rate fraction),
$\Lout$ (outflow power),
$\tout$ (outflow expansion time),
$\Eout$ (total outflow energy),
$\vout$ (outflow velocity),
$\trej$ (rejuvenation time),
$\delta$ (duty cycle),
and $\dotMgrow$ (net sMBH growth rate).
The SMBH mass and accretion rate are fixed at $\Mp=10^8\Msun$ and $\dotMp=1$.}
\label{fig:Bondi}
\end{figure*}

The hyper-Eddington accretion develops outflows, with the kinetic power given by \citep{Liu2024}
\begin{equation}
\label{eq:Lout}
\Lout=\eta_{\rm rad}\facc (1-f_{\rm adv}) \dot{M}_{\rm s} c^2,
\end{equation}
where $\eta_{\rm rad}\approx 0.1$ denotes the radiative efficiency,
and $f_{\rm adv}\approx 0.9$ is the advection fraction.
The value of $f_{\rm adv}$ is quite uncertain due to the complex interplay of advection, photon trapping, and outflow dynamics \citep[e.g., simulations in][]{Takeuchi2009}.
The factor $\facc =\left(X_{\rm in}/{\rout}\right)^s$ represents the fraction of the Bondi accretion rate that ultimately reaches the sMBH,
{where $\rout=\min\{\Xout,\rtr\}$ is the injection radius of the outflows,
$\rtr=3\dotMs h \rg$ is the photon trapping radius \citep[e.g.,][]{Kato2008,Kitaki2021},
$h\sim 0.45$ is the aspect ratio of sMBH disk \citep[see Equation~(A14) in][]{Liu2024},
and $\rg=G\Ms/c^2$ is the gravitational radius of the sMBH.}
% \citep[see Equation~(A2) and accompanying explanations in][]{Liu2024}.
Here, the index $s$, ranging from 0 to 1, depends on the ratio of the radiation energy to the dissipated gravitational energy \citep{Begelman2012}.
{A moderate index of $s=1/2$ is adopted in this work.}
The inner radius of the sMBH disk is taken as $X_{\rm in}=6\,\rg$, corresponding to the innermost stable circular orbit around a Schwarzschild black hole.
The outer radius $\Xout=(\Delta R)^4\Mp/4 R^3\Ms$ is derived from angular momentum conservation of the infalling material \citep[see Equation~(A1) in][]{Liu2024},
where $\Delta R=\min\{\RBon, \RHill\}$ is the characteristic size of the accretion region.
As shown in Figure~\ref{fig:HR}, $\Xout$ is typically an order of magnitude smaller than $\RHill$, except for the low-mass sMBHs in the inner disk.
When $\RHill<\RBon$, we have $\Delta R=\RHill$, yielding $\Xout = \RHill/12$.
This indicates that the sMBH disk remains gravitationally bound to the sMBH and is not tidally disrupted by the central SMBH.
As illustrated in Figure~\ref{fig:Bondi}, $\Lout$ increases with sMBH mass $\Ms$ in the inner disk.
However, in the outer disk ($R/\Rg\gtrsim 100$), $\Lout$ exhibits a non-monotonic dependence on $\Ms$: it rises as $\Ms$ increases from $10$ to $100\,\Msun$, but declines when $\Ms$ increases further to $10^3\,\Msun$.
This behavior results from the suppression of spherical accretion deviation due to increased differential motion.
Additionally, $\Lout$ decreases with disk radius $R$, primarily due to the rapid decline in gas density $\rho$.

Shocks within the Bondi radius region heat the gas and temporarily halt accretion over an accretion timescale,
which corresponds to the outflow expansion timescale $\tout$.
{Based on the standard cavity evolution model given by \citet[Equation~(21)]{Weaver1977},
$R(t) = (125/154\pi)^{1/5} \Lout^{1/5} \rho^{-1/5}  t^{3/5}$,
the timescale $\tout$ is derived as follows:}
\begin{equation}
\label{eq:toutEner}
\tout = \left(\frac{{231}\Mout \Rout^2}{{250\pi^2}\Lout}\right)^{1/3},
\end{equation}
where $\Rout=\min\{\RBon, \RHill, H\}$ is the radius at which infalling gas from the SMBH disk is halted by the outflow,
and the mass enclosed within $\Rout$ is $\Mout=4\pi \Rout^3\rho/3$.
{The outflow can break out of the disk under certain conditions \citep{Chen2023}; however, here we focus on its interaction with the AGN disk.
The resulting turbulence is primarily constrained by $\Rout$ provided the overall disk structure remains intact despite the outflow.
When $\Rout=H$, this corresponds to the maximum lateral extent of the cavity, as the outflow expands vertically but not horizontally after emerging from the disk (see upper left panel of Figure~\ref{fig:cartoon}). 
If $\Rout=\RBon$ or $\RHill$, it represents the “choking” radius. Additionally, the timescale $\tout$ in Equation~(\ref{eq:toutEner}) differs slightly from the order-of-magnitude estimate given by \citet[Equation~(9)]{Liu2024} by a factor of $(231/125\pi^2)^{1/3}$.}
As shown in Figure~\ref{fig:Bondi}, $\tout$ increases from hours to months as the sMBH location varies from $10$ to $10^4 \Rg$.
This trend reflects the longer time required to accumulate sufficient energy to break out of the disk at larger radii, despite the lower gas density.
% The accumulated kinetic energy within this region is
The kinetic energy accumulated within the region is given by
\begin{equation}
\Eout={\frac{6}{11}} \Lout \tout,
\end{equation}
{where the factor $6/11$ arises because the total injected energy is partitioned between the kinetic energy of the outflow and the internal energy of the cavity \citep[see Equation~(20) in][]{Weaver1977}.}
As indicated in Figure~\ref{fig:Bondi}, $\Eout$ ranges from $10^{44}$ to $10^{49} \rm erg$.
In the inner disk region, $\Eout$ increases noticeably with sMBH mass, consistent with stronger accretion at higher masses.
In the outer disk, however, $\Eout$ depends only weakly on $\Ms$ but rises significantly with $R$.
This behavior arises because the greater disk height at larger radii, together with the longer duration of Bondi accretion ($\tout$), allows more energy to accumulate before the outflow breaks out.
The outflow velocity is given by
\begin{equation}
\vout =  \frac{{3}\Rout}{{5}\tout},
\end{equation}
{where the factor $3/5$ accounts for the non-constant expansion speed of the outflow \citep[see Equation~(21) in][]{Weaver1977}.}

After a Bondi explosion, strong outflows from a Type I AMS collide with AGN disk gas, generating shocks that fragment into small-scale turbulent vortices.
The resulting turbulent viscous stress transports angular momentum outward from the inner disk regions.
This mechanism is analogous to the case of a supernova exploding in an AGN disk \citep{Rozyczka1995, Moranchel2021}.
The momentum carried by the outflows is given by $P=\sqrt{2\Mout\Eout}$.
As shown in Figures 3-5 of \cite{Rozyczka1995} and Figure 5 of \cite{Moranchel2021},
the shocked shell is advected inward or outward over a distance of order $\Rout$ on the timescale of one or several orbital periods after the Bondi explosion.
This process redistributes angular momentum in the disk gas through shock interaction and angular momentum exchange.
When the Bondi explosion shocks interact with the disk gas, disk fluid elements facing the inner shock lose angular momentum,
while those facing the outer shock gain angular momentum. 
The net effect is a radial transport of angular momentum from the inner disk to the outer disk.
The total angular momentum transferred is approximately
\begin{equation}\label{eq:J}
J=\displaystyle\frac{\mu}{\pi}P\Rout,
\end{equation}
where fudge factor $\mu$, expected to be of order unity, can be calibrated from numerical simulations \citep{Moranchel2021}.
For simplicity, we adopt $\mu=1$ in the following calculations.
The time-averaged angular momentum transfer rate per sMBH is
\begin{equation}\label{eq:dotJA}
\dot{J}=\frac{J}{\trej+\tout},
\end{equation}
where $\trej=\Rout/\alpha\cs$ is the rejuvenation time,
during which the gas refills the cavity \citep{Wang2023SgrA,Liu2024}.
As shown in Figure~\ref{fig:Bondi}, $\trej$ increases from days to $100$ years with {increasing} $R$, significantly longer than $\tout$ (hours to months).

The powerful outflows quench the hyper-Eddington accretion, thereby preventing runaway growth of the sMBH.
The time-averaged dimensionless mass growth rate (net accretion rate) is defined as $\dotMgrow\equiv \facc\delta\dotMs$,
where $\delta \equiv \tout/(\tout+\trej)$ is the duty cycle.
As shown in Figure~\ref{fig:Bondi},
the duty cycle $\delta \sim 0.004-0.016$ indicates that most AMSs are in the rejuvenation phase.
During this phase, the accretion rate onto a Type II AMS drops significantly, and the sMBH can no longer efficiently accrete the hot, low-density gas inside the cavity.
In the outer disk region, $\dotMgrow$ decreases with increasing sMBH mass $\Ms$,
which slows logarithmic mass growth and prevents disruptive rapid accretion.
This is consistent with the analytically derived average growth timescale presented in \cite{Liu2024}.

\subsection{Scenario B: steady accretion of Eddington limit}
\label{sect:steady}

In Scenario B, we adopt the optically thick wind model \citep{Meier2012,Zhou2019},
in which the outflow and inflow maintain a steady state.
The wind is launched from the inner boundary and undergoes acceleration until reaching the critical radius.  
% at $r_{\rm i}=\dot{M}_{\rm s}r_{\rm isco}$
The kinetic power of the outflow is limited by the Eddington luminosity $ \Lout= L_{\rm Edd,s}$.
As illustrated in Figure~\ref{fig:HR}, the Hill radius $\RHill$ generally exceeds the disk height $H$,
allowing the outflow to break out of the SMBH disk.
The outflow velocity at $\RHill$ can be derived from the energy equation $4\pi \Rout^2 \vout^3 \rho=\Lout$,
where $\Rout=\min\{\RHill, H\}$ is the outflow radius
and $\RBon$ is given by Equation~(\ref{eq:RBon}).
This yields
\begin{equation}\label{eq:vout}
\vout=\left(\frac{\Lout}{4\pi\Rout^2\rho}\right)^{1/3}.
\end{equation}
The momentum flux carried by the outflow is $\dot{P}=4\pi \Rout^2\vout^2\rho$.
Similar {to} Equation~(\ref{eq:J}), the angular momentum transfer rate is $\dot{J}=\mu\dot{P}\Rout/\pi$.
Substituting Equation~(\ref{eq:vout}), we obtain 
\begin{equation}\label{eq:dotJB}
\dot{J}={\left(\frac{2}{\pi}\right)^{2/3}}\mu\Lout^{2/3}\Rout^{5/3}\rho^{1/3}.
\end{equation}
Although this formulation provides a first-order approximation, it captures the essential physics of the angular momentum transfer process.
High-resolution hydrodynamic simulations will be crucial to directly model the interaction between AMS outflows and the disk gas \citep[e.g.,][]{Zhang2025Lizhong},
from which the value of $\mu$ is derived.
Such simulations help clarify the microscopic transition from shock to turbulence and thereby establish a more robust physical foundation for the torque model described by $\dot{J}$.

\section{angular momentum transfer}
\label{sect:AMtransfer}

\subsection{sMBH mass function and location}
\label{sect:NUM}

The sMBH mass function remains poorly constrained.
Observations of BBH mergers by LIGO/Virgo suggest a mass distribution of the form $dN/d\Ms\varpropto \Ms^{-\beta}$, with a power-law index of $\beta=1.3_{-1.7}^{+1.4}$ \citep{Abbott2019}.
Although it is uncertain whether these mergers occur in AGN disks, the observed index may overestimate the intrinsic steepness of the mass function,
since mergers involving more massive progenitors are more easily detected.
In dense gaseous environments such as AGN disks, accretion is expected to harden the mass function of stars \citep{Wang2023,Chen2024} and black holes, shifting the index by $\Delta \beta \sim 1.3$ \citep[][]{Yang2019}.
A similar hardening effect can also result from BBH merger processes \citep{Li2025},
with the maximum sMBH mass primarily limited by the AGN disk lifetime \citep{Xue2025}.
To cover a plausible range of mass function shapes, we consider $\beta=1,\,2$, and $3$ in our calculations, along with a minimum sMBH mass $M_{\rm min}=10\Msun$ and a maximum sMBH mass $M_{\rm max}=10^3\Msun$.

sMBHs embedded in an AGN disk inevitably undergo orbital migration due to dynamical friction with the surrounding gas.
For simplicity, we assume their spatial distribution follows a power-law form, $\rmd N/\rmd R \propto R^{\gamma}$, with inner and outer disk boundaries set to $\Rmin=10\Rg$ and $\Rmax=10^4\Rg$, respectively.
A value of $\gamma=1$ corresponds to a uniform surface number density.
In an $\alpha$-disk, conservative inward migration of sMBHs would yield $\gamma=1/4\,(2/5$, or $5/2)$ in the outer, middle, or inner regions, respectively.
When BBH mergers during migration are considered, $\gamma$ is expected to be smaller.
To cover a reasonable parameter space, we consider $\gamma = -1, 0, 1$, and $2$.
The joint mass and radial distribution of sMBHs is described by
\begin{equation}\label{eq:N}
\mathcal{N}\equiv\frac{\rmd N}{\rmd M \rmd R}=\mathcal{N}_0 M^{-\beta}R^{\gamma},
\end{equation}
where $\mathcal{N}_0$ is the normalization constant.

The total gas mass in the SMBH disk is $M_{\rm disk}=\int_{\Rmin}^{\Rmax}4\pi R\rho H \rmd R=2.86\times10^6\,\Msun$. 
The total sMBH mass, $\MTot$, is uncertain.
We parameterize it as $\MTot=\zeta M_{\rm disk}={2.86\times10^4\zeta_{\bar{2}}}M_{\rm disk,6}\Msun$,
where {$\zeta_{\bar{2}}=\zeta/10^{-2}$}, and adopt $\zeta={10^{-2}}$ as a reference value, implying that sMBHs comprise about {1\%} of the disk mass.
The normalization $\mathcal{N}_0$ is then derived from the condition $\MTot=\int_{\Rmin}^{\Rmax} \mathcal{N} M \rmd M \rmd R$.
We emphasize that ${\zeta_{\bar{2}}}$ here is quite uncertain.
\cite{Chen2024} estimated a population of about $10^5$ massive stars in the outer unstable region of an AGN disk around a $10^8\,\Msun$ SMBH, considering accretion efficiency and thermal feedback.
In comparison, this study adopts a conservative, lower number of sMBHs to estimate the viscosity contributed by AMSs.
In reality, accretion and outflows from ordinary stars, neutron stars, and white dwarfs may also generate turbulence and enhance disk viscosity, but these are beyond the scope of this paper and left for future study.

The spatial distribution of sMBHs in AGN disks is not yet well constrained.
Potential migration traps may occur at various locations: in the inner disk, at the transition radius between sub- and super-Keplerian rotation \citep{Peng2021};
in the middle region, due to gravitational torques from the SMBH disk \citep{Bellovary2016};
or in the outer disk, through so-called “thermal torques” \citep{Grishin2024, Vaccaro2025}.
These traps are considered likely sites for BBH mergers \citep{Vaccaro2025b, Samsing2025}.
On the other hand, merged sMBHs may subsequently escape from such traps due to GW recoil kicks or gap-opening mechanisms \citep{Gilbaum2025}.
In this work, we focus primarily on the general role played by AMSs in driving viscosity in AGN disks, and do not address the detailed effects of migration trapping,
which we leave for future investigation.

\begin{figure*}
\centering
\includegraphics[width=\linewidth]{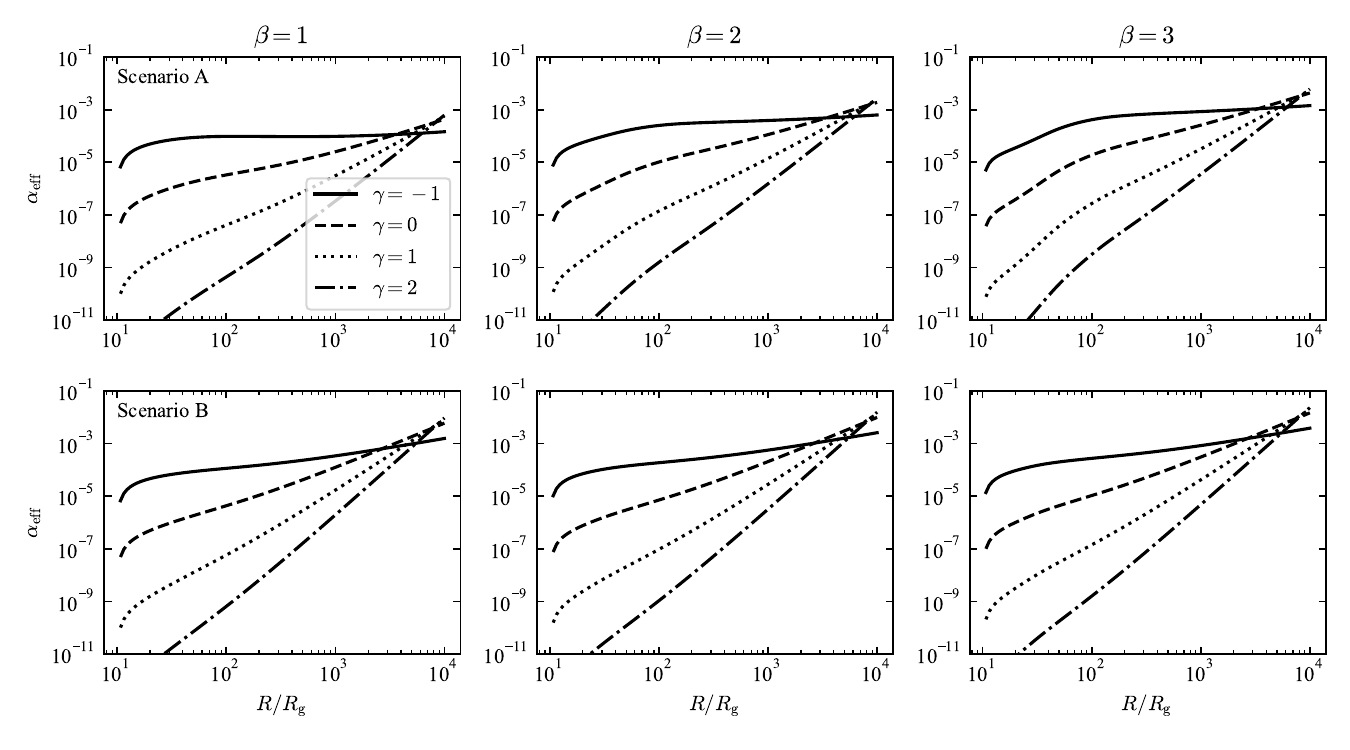}
\caption{Radial profiles of the effective viscosity parameter $\alphaeff$ for Scenario A (top) and B (bottom),
comparing different sMBH number density distributions parameterized by $\beta$ and $\gamma$.
The SMBH mass and accretion rate are fixed at $\Mp=10^8\Msun$ and $\dotMp=1$.}
\label{fig:alphaeff}
\end{figure*}

\subsection{effective viscosity parameter}
\label{sect:result}

Using Equation~(\ref{eq:dotJA}) in Scenario A, Equation~(\ref{eq:dotJB}) in Scenario B, and the sMBH distribution given in Equation~(\ref{eq:N}),
we derive the differential angular momentum transfer rate contributed by the entire sMBH population as a function of radius $R$
\begin{equation}\label{eq:dotJ}
\dotJ=
\int_{M_{\rm min}}^{M_{\rm max}}
\dot{J}\mathcal{N}
\rmd M.
\end{equation}
As shown in Figure~\ref{fig:Bondi},
the rejuvenation timescale $\trej$ ranges from $1-10^4\rm yr$, depending only weakly on the sMBH mass $\Ms$ but strongly on radius $R$.
Since $\tout\ll\trej$, the duty cycle $\delta$ is very small.
The total viscous torque in an $\alpha$-disk is $T=2\pi R^2 \int t_{r\phi} dz=4\pi R^2H\alphaeff p$,
where the standard viscosity prescription $t_{r\phi}=\alphaeff p$ is adopted \citep{Kato2008}.
Applying the relation $\dotJ=\rmd T/\rmd R$, the effective viscosity parameter can be expressed as
\begin{equation}\label{eq:alphaeff}
\alphaeff=\frac{1}{4\pi R^2H p}\int_{\Rmin}^R \dotJ(R^\prime) \rmd R^\prime.
\end{equation}
The effective viscosity $\alphaeff$ is directly proportional to the total sMBH mass, since $\alphaeff\propto\mathcal{N}_0\propto \MTot\propto\zeta$.

In Figure~\ref{fig:alphaeff}, we {compare} Scenario A (Bondi explosions, top) and Scenario B (steady Eddington accretion, bottom), showing the variation of effective viscosity $\alphaeff$ with $R$ for different sMBH mass function exponents ($\beta=1$ for a flat mass function, $\beta=3$ for a steep mass function) and radial distribution exponents ($\gamma=-1$ for inner-disk concentration, $\gamma=2$ for outer-disk concentration).
In Scenario A, $\alphaeff$ ranges from ${10^{-11}}$ to ${10^{-2}}$ from inner to outer regions.
$\alphaeff$ is small in the inner disk ($R/\Rg < 100$, ${10^{-11}-10^{-4}}$) and increases to ${10^{-6}-10^{-2}}$ in the outer disk ($R/\Rg > 10^3$).
For steeper mass function ($\beta=3$) in the outer region ($R/\Rg=10^4$), there is more low-mass sMBHs, generating more energy and turbulence consequently and leading to a larger $\alphaeff\sim {10^{-2}}$.
Outer-disk concentration ($\gamma=2$) increases $\alphaeff$ in the outer disk because concentrated sMBHs inject more energy, boosting angular momentum transfer efficiency.
For inner-disk concentration ($\gamma=-1$), $\alphaeff$ is slightly dependent on the location $R$.
In Scenario B, $\alphaeff$ is generally slightly higher than in Scenario A.
Steady outflows continuously inject energy, maintaining more stable turbulence and higher angular momentum transfer efficiency (larger $\dot{P}$ in Equation~(\ref{eq:dotJB}), leading to larger $\dot{J}$).
$\alphaeff$ has a similar dependence on the mass function and the position distribution as Scenario A.
In total, $\alphaeff$ in both scenarios covers the viscosity range required for AGN disks {if a larger $\zeta$ value is adopted}.
This confirms AMS as a critical viscosity source alongside MRI and GI.

\begin{deluxetable}{lcccc ccccc}
\footnotesize
\tablecaption{The global effective viscosity $\alphaAMS$ for $\Mp=10^8\,\Msun$ and $\dotMp=1$.}
\label{tab:alphaAMS}
\tablehead{
& \multicolumn{3}{c}{Scenario A} && \multicolumn{3}{c}{Scenario B} \\
\cline{2-4} \cline{6-8}
& \colhead{$\beta=1$} & \colhead{$\beta=2$}& \colhead{$\beta=3$}
&& \colhead{$\beta=1$} & \colhead{$\beta=2$}& \colhead{$\beta=3$}}
\startdata
$\gamma=-1$&$0.0003$&$0.0011$&$0.0025$&&$0.0027$&$0.0045$&$0.0067$&\\
$\gamma=0$&$0.0008$&$0.0033$&$0.0075$&&$0.0104$&$0.0173$&$0.0257$&\\
$\gamma=1$&$0.0010$&$0.0042$&$0.0096$&&$0.0144$&$0.0238$&$0.0354$&\\
$\gamma=2$&$0.0011$&$0.0047$&$0.0106$&&$0.0165$&$0.0272$&$0.0405$&\\
\enddata
\end{deluxetable}

\begin{figure*}
\centering
\includegraphics[width=\linewidth]{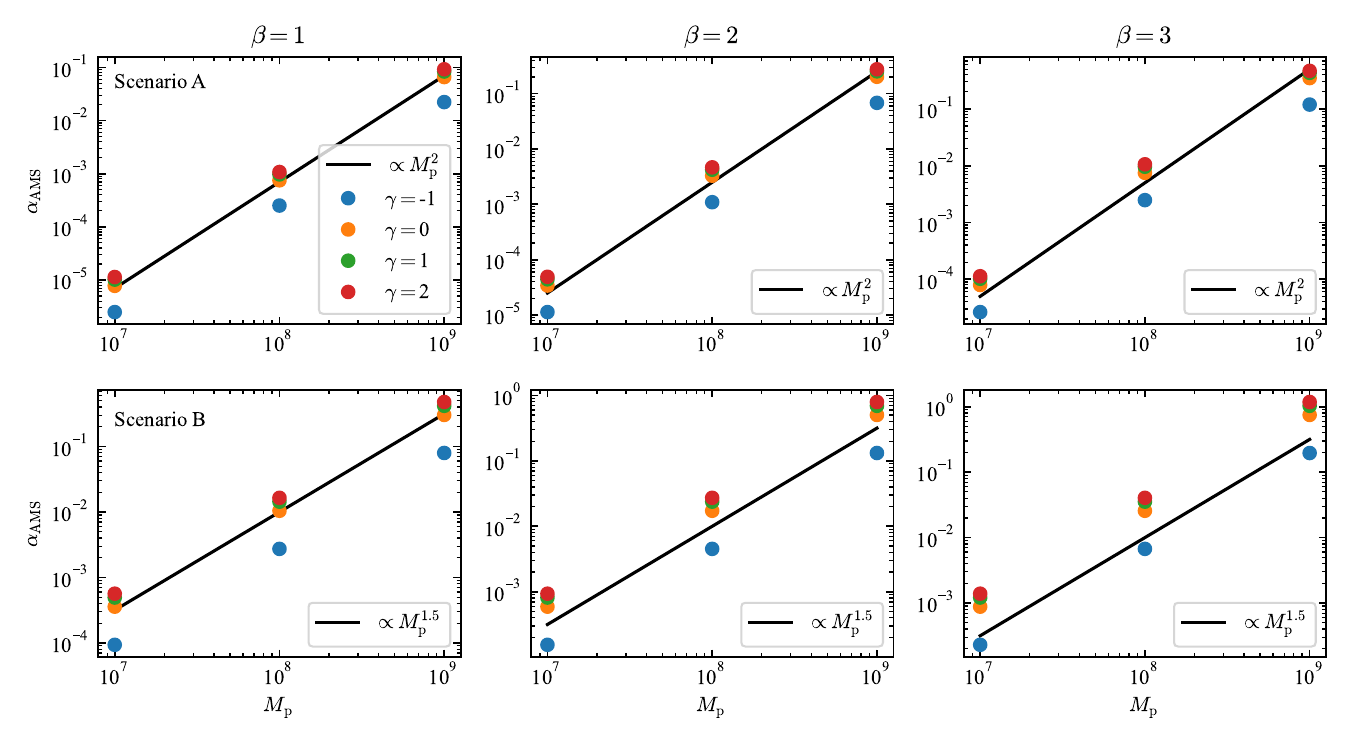}
\caption{Global effective viscosity $\alphaAMS$ as a function of $\Mp$ for Scenario A (top) and B (bottom).
The cases for different sMBH number density distributions in the parameter space of $\beta$ and $\gamma$ are compared.
Corresponding numerical values are listed in Tables~\ref{tab:alphaAMS} and \ref{tab:alpha_M7_9}.}
\label{fig:alphaAMS_M}
\end{figure*}

\begin{figure*}
\centering
\includegraphics[width=\linewidth]{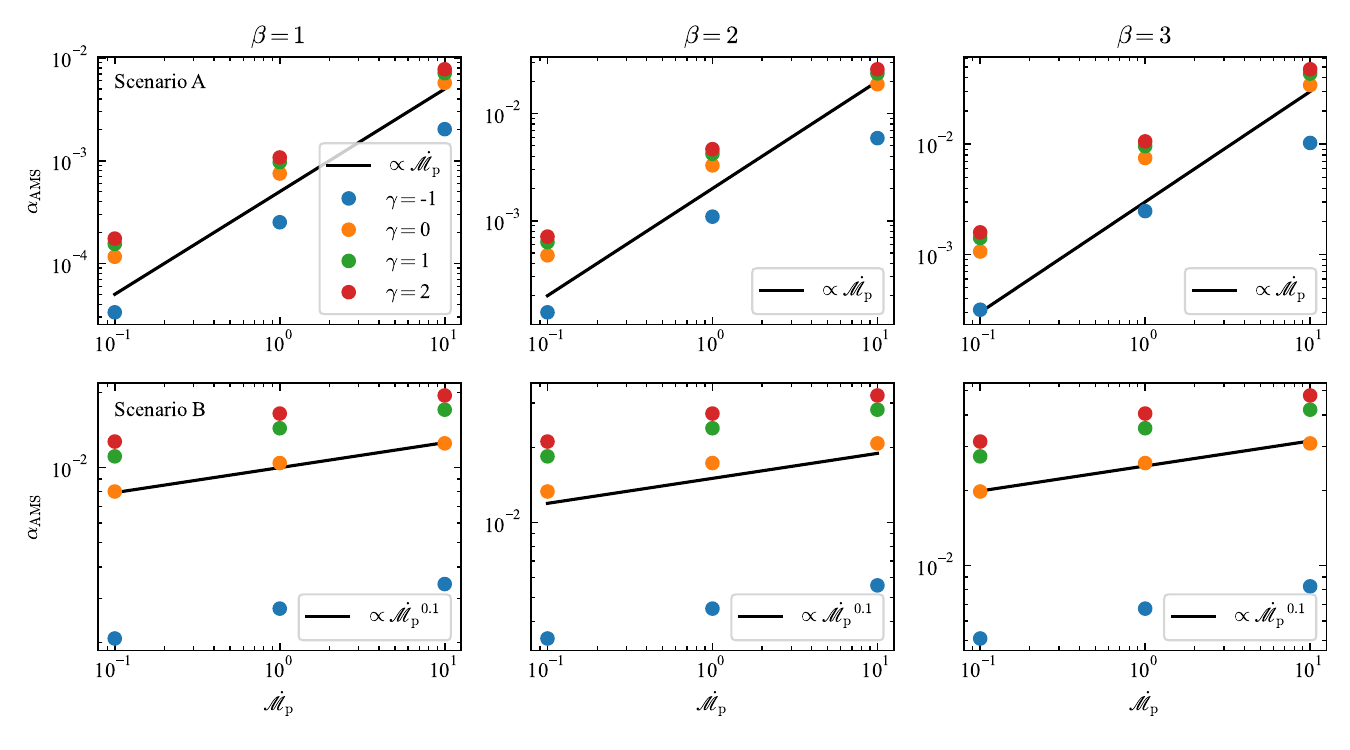}
\caption{Global effective viscosity $\alphaAMS$  as a function of $\dotMp$ for Scenario A (top) and B (bottom).
The cases for different sMBH number density distributions in the parameter space of $\beta$ and $\gamma$ are compared.
Corresponding numerical values are listed in Tables~\ref{tab:alphaAMS} and \ref{tab:alpha_dotM01_10}.}
\label{fig:alphaAMS_dotM}
\end{figure*}

% The global effective viscosity parameter is obtained by integrating Equation~(\ref{eq:dotJ}) over radius $R$.
The total angular momentum transfer rate from the sMBH population is given by $\dotJ_{\rm tot}=\int_{\Rmin}^{\Rmax} \dotJ\rmd R$,
while the total angular momentum of the SMBH disk is $\mathcal{J}_{\rm tot}=\int_{\Rmin}^{\Rmax} 2\pi R^3\Sigma\Omega \rmd R$.
The migration timescale driven by AMS feedback is then $\tmig=\mathcal{J}_{\rm tot}/\dotJ_{\rm tot}$.
For a standard $\alpha$ disk, the viscosity timescale is $\tvis=(R/H)^2/\alpha \OmgK$ \citep{Kato2008}.
Equating these timescales yields the global effective viscosity parameter as
\begin{equation}
\alphaAMS\equiv\frac{\dotJ_{\rm Tot}}{\Omega_{\rm m} \mathcal{J}_{\rm Tot}}\left(\frac{\Rmax}{H_{\rm max}}\right)^2,
\end{equation}
where $\Omega_{\rm m},H_{\rm max}$ is the angular velocity and SMBH disk height at radius $\Rmax$, respectively.
The resulting values of $\alphaAMS$ are listed in Table~\ref{tab:alphaAMS}, ranging from ${3\times10^{-4}}$ to ${10^{-2}}$ for Scenario A and from ${2.7\times10^{-3}}$ to ${4.1\times10^{-2}}$ for Scenario B.
Additionally, Table~\ref{tab:alpha_M7_9} provides $\alphaAMS$ for different SMBH mass ($\Mp=10^7$ and $10^9\,\Msun$), with the corresponding results plotted in Figure~\ref{fig:alphaAMS_M}.
The results are consistent with scaling relations of $\alphaAMS\propto\Mp^2$ for Scenario A and $\alphaAMS\propto\Mp^{1.5}$ for Scenario B.
Similarly, Table~\ref{tab:alpha_dotM01_10} lists $\alphaAMS$ for different SMBH accretion rate ($\dotMp=0.1$ and $10$), as shown in Figure~\ref{fig:alphaAMS_dotM}.
These {follow} the relations $\alphaAMS\propto\dotMp$ for Scenario A and $\alphaAMS\propto\dotMp^{0.1}$ for Scenario B.
The above scaling relations are primarily determined by variations of the disk height $H$ and the Hill radius with $\Mp$ and $\dotMp$.
Further details of the radial profiles are provided in Appendix~\ref{appdix}.
{It is noteworthy that $\alphaAMS$ exceeds 0.1 in most cases for $\Mp=10^9\Msun$,
where a self-consistent solution of AMS-AGN disk becomes necessary.
But this extreme scenario lies beyond the scope of the present study.
Through multiple iterations, a value of $\alphaAMS\sim 0.1-1$ can be preliminarily anticipated.
A more detailed investigation, particularly for systems with higher SMBH and sMBH masses, remains a subject for future work.}

\begin{deluxetable}{lcccc ccccc ccccc ccccc}
\footnotesize
\tablecaption{The same as Table~\ref{tab:alphaAMS} but for $\Mp=10^7\Msun$ and $\Mp=10^9\Msun$.}
\label{tab:alpha_M7_9}
\tablehead{
& \multicolumn{7}{c}{$\Mp=10^7\Msun,~\dotMp=1$} && \multicolumn{7}{c}{$\Mp=10^9\Msun,~\dotMp=1$} \\
\cline{2-8} \cline{10-16}
& \multicolumn{3}{c}{Scenario A} && \multicolumn{3}{c}{Scenario B} 
&& \multicolumn{3}{c}{Scenario A} && \multicolumn{3}{c}{Scenario B} \\
\cline{2-4} \cline{6-8} \cline{10-12} \cline{14-16}
& \colhead{$\beta=1$} & \colhead{$\beta=2$}& \colhead{$\beta=3$} 
&& \colhead{$\beta=1$} & \colhead{$\beta=2$}& \colhead{$\beta=3$}
&& \colhead{$\beta=1$} & \colhead{$\beta=2$}& \colhead{$\beta=3$} 
&& \colhead{$\beta=1$} & \colhead{$\beta=2$}& \colhead{$\beta=3$}}
\startdata
$\gamma=-1$&$0.000002$&$0.000011$&$0.000026$&&$0.00009$&$0.00015$&$0.00023$&&$0.022$&$0.068$&$0.119$&&$0.08$&$0.13$&$0.20$&\\
$\gamma=0$&$0.000008$&$0.000034$&$0.000079$&&$0.00036$&$0.00059$&$0.00088$&&$0.066$&$0.201$&$0.350$&&$0.30$&$0.50$&$0.75$&\\
$\gamma=1$&$0.000010$&$0.000044$&$0.000102$&&$0.00049$&$0.00082$&$0.00121$&&$0.084$&$0.252$&$0.432$&&$0.42$&$0.69$&$1.03$&\\
$\gamma=2$&$0.000011$&$0.000049$&$0.000113$&&$0.00056$&$0.00093$&$0.00139$&&$0.092$&$0.275$&$0.469$&&$0.48$&$0.79$&$1.18$&\\
\enddata
\end{deluxetable}

\begin{deluxetable}{lcccc ccccc ccccc ccccc}
\footnotesize
\tablecaption{The same as Table~\ref{tab:alphaAMS} but for $\dotMp=0.1$ and $\dotMp=10$.}
\label{tab:alpha_dotM01_10}
\tablehead{
& \multicolumn{7}{c}{$\Mp=10^8\Msun,~\dotMp=0.1$} && \multicolumn{7}{c}{$\Mp=10^8\Msun,~\dotMp=10$} \\
\cline{2-8} \cline{10-16}
& \multicolumn{3}{c}{Scenario A} && \multicolumn{3}{c}{Scenario B} 
&& \multicolumn{3}{c}{Scenario A} && \multicolumn{3}{c}{Scenario B} \\
\cline{2-4} \cline{6-8} \cline{10-12} \cline{14-16}
& \colhead{$\beta=1$} & \colhead{$\beta=2$}& \colhead{$\beta=3$} 
&& \colhead{$\beta=1$} & \colhead{$\beta=2$}& \colhead{$\beta=3$}
&& \colhead{$\beta=1$} & \colhead{$\beta=2$}& \colhead{$\beta=3$} 
&& \colhead{$\beta=1$} & \colhead{$\beta=2$}& \colhead{$\beta=3$}}
\startdata
$\gamma=-1$&$0.00003$&$0.00014$&$0.00032$&&$0.0021$&$0.0034$&$0.0051$&&$0.002$&$0.006$&$0.010$&&$0.003$&$0.006$&$0.008$&\\
$\gamma=0$&$0.00012$&$0.00048$&$0.00107$&&$0.0080$&$0.0133$&$0.0198$&&$0.006$&$0.019$&$0.034$&&$0.013$&$0.021$&$0.031$&\\
$\gamma=1$&$0.00015$&$0.00063$&$0.00141$&&$0.0111$&$0.0184$&$0.0273$&&$0.007$&$0.024$&$0.044$&&$0.017$&$0.028$&$0.042$&\\
$\gamma=2$&$0.00017$&$0.00071$&$0.00159$&&$0.0127$&$0.0211$&$0.0313$&&$0.008$&$0.026$&$0.048$&&$0.019$&$0.032$&$0.048$&\\
\enddata
\end{deluxetable}

\section{conclusion and discussion}
\label{sect:discussion}

We studied {AMS} dynamical evolution and viscosity contributions in AGN disks.
Two kinds of outflows, episodic outflows driven by Bondi explosion (Scenario A) and steady outflows under Eddington limit (Scenario B),
lead to turbulence through the shock interaction with the AGN disk gas.
This turbulence facilitates outward angular momentum transport in the disk.
By comparing the migration timescale with the viscous timescale, we computed the global effective viscosity parameter $\alphaAMS$,
which ranges from ${3\times10^{-4}}$ to ${10^{-2}}$ for Scenario A and from ${2.7\times10^{-3}}$ to ${4.1\times10^{-2}}$ for Scenario B.
In most sMBH spatial distributions, the viscosity contributed by sMBHs is higher in the outer disk regions and lower in the inner regions.

For Scenario A, AMSs exhibit biphasic evolution, alternating between Type I (Bondi explosion, $\tout$: hours to months) and Type II (rejuvenation, $\trej$: ${10^{-2}-10^2}$ years) state, with a duty cycle of $\delta \approx {0.002-0.009}$.
This process results in a net dimensionless sMBH growth rate of $\dotMgrow = {1-10^{3}}$,
which provides seed black holes for gravitational wave (GW) events with large masses while preventing excessive sMBH growth.
The sMBH accretion regions are confined by the disk height $H$ and the Hill radius $\RHill$, where $H <\RBon$ and $\RHill < \RBon$ in most regions.
In the outer disk, $\Delta v >\cs$ suppresses spherical accretion, thereby reducing the sMBH accretion rate $\dotMs$.
Both Scenario A and Scenario B yield $\alphaAMS$ values covering the required AGN disk viscosity range.
The effective viscosity in the outer disk is 2-4 orders of magnitude higher than in the inner disk, indicating that AMS dominates viscosity in the outer regions.
A steep sMBH mass function ($\beta=3$) combined with outer-disk concentration ($\gamma=2$) can raise the effective viscosity to $\alphaeff\gtrsim0.1$ in the outer disk.
Conversely, a flat mass function ($\beta=1$) and inner-disk concentration ($\gamma=-1$) lead to lower $\alphaeff$ values.
This confirms that the sMBH population distribution is a key regulator of viscosity in AGN disks.

The present model assumes power-law distributions for the sMBH population: $\rmd N/\rmd \Ms \propto \Ms^{-\beta}$ in mass and $\rmd N/\rmd R \propto R^\gamma$ in radius.
However, ``migration traps", such as the transition zone between sub- and super-Keplerian rotation in the inner disk \citep{Peng2021}, or regions dominated by thermal torques in the outer disk \citep{Grishin2024}, 
are known to exist in AGN disks and may cause sMBHs to accumulate at specific radii \citep[e.g., $R \approx 10^4 \Rg$,][]{Trani2025}.
Future high-resolution $N$-body simulations are needed to refine the distribution function,
and are expected to shift the peak location of $\alphaeff$ seen in Figure~\ref{fig:alphaeff}.
Gravitational-wave observations will provide key constraints on these model parameters \citep[e.g.,][]{Ford2025,McKernan2025}.
While ground-based detectors LIGO/Virgo can search for binary black hole (BBH) mergers in AGN disks,
the mHz band accessible to space-based detectors like Laser Interferometer Space Antenna \citep[LISA,][]{Babak2017}, Taiji \citep{Hu2017}, and Tianqin \citep{Luo2016} will be particularly suited to detecting EMRIs \citep{Wang2023SgrA} and intermediate-mass binary black hole merger events \citep{Dutta2025}.
These multi-band GW observations will help determine the values of $\mathcal{N}_0$, $\beta$, and $\gamma$, offering an empirical test for the theoretical predictions of $\alphaeff$.

\begin{acknowledgments}
We thank Yi-Lin Wang for useful discussions and the anonymous referee for useful comments.
We acknowledge funding support from the National Natural Science Foundation of China under the grant Nos.\ 12333003, 12025301, the National Key R\&D Program of China (2021YFA1600404), and the Strategic Priority Research Program of the Chinese Academy of Sciences.
This work is supported by the China Postdoctoral Science Foundation under Grant Number 2025M783227.
\end{acknowledgments}

\appendix

\restartappendixnumbering
\section{Effective Viscosity vs. SMBH Mass and Accretion Rate}
\label{appdix}

\begin{figure*}
\centering
\includegraphics[width=\linewidth]{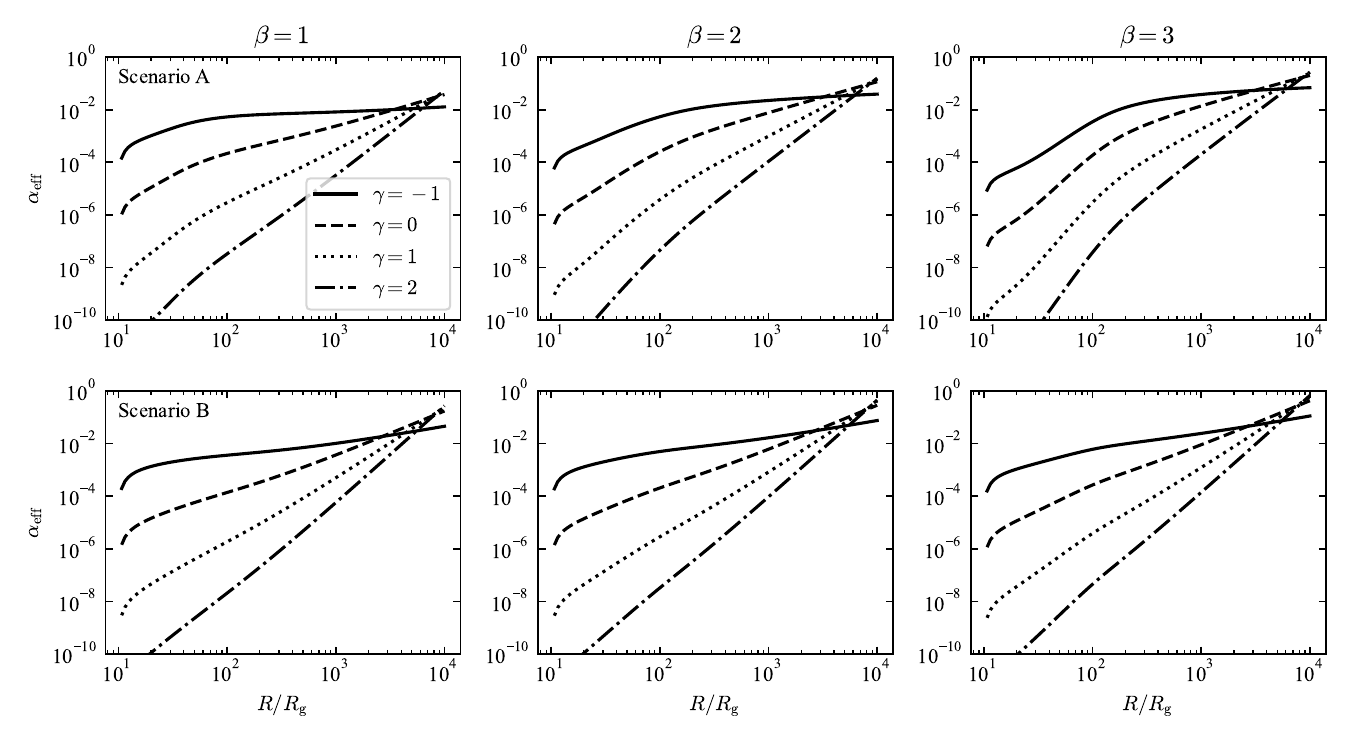}
\caption{The same as Figure~\ref{fig:alphaeff} but for $\Mp=10^9\Msun$.}
\label{fig:alphaeff_M9}
\end{figure*}

\begin{figure*}
\centering
\includegraphics[width=\linewidth]{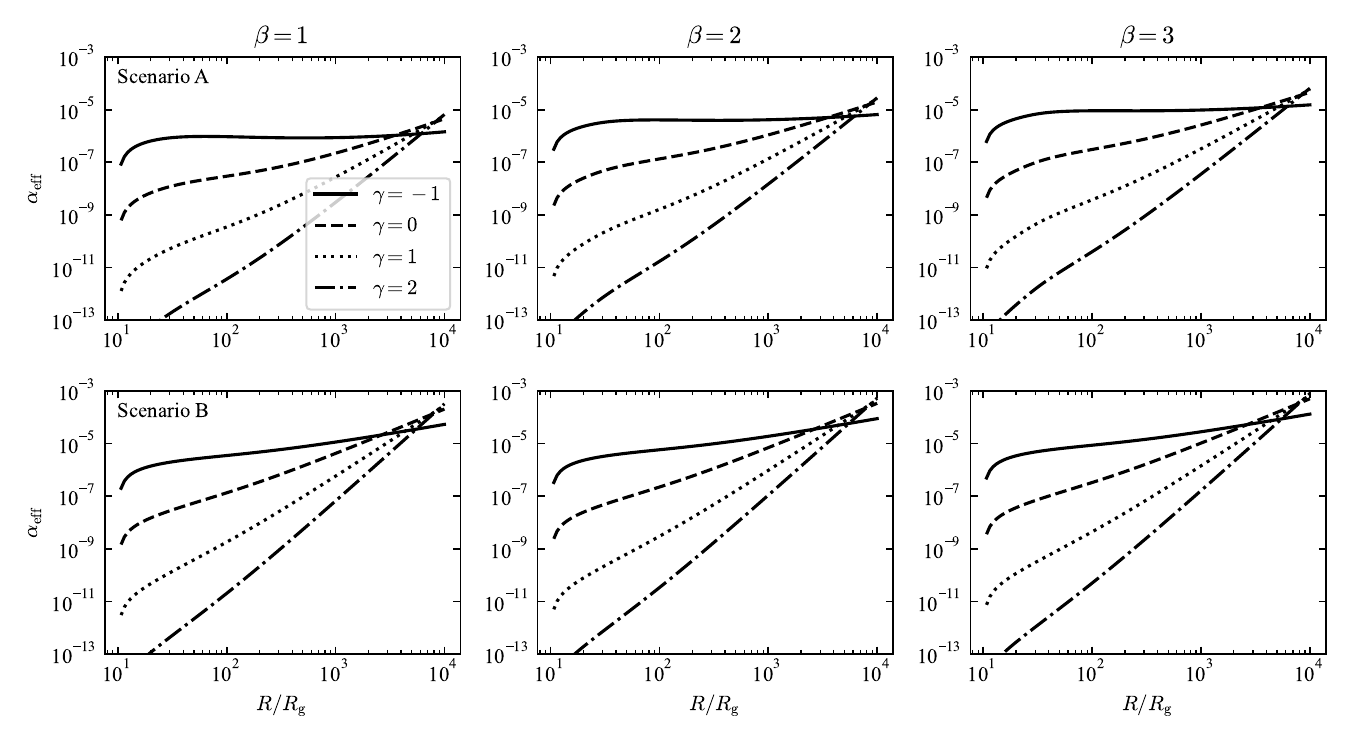}
\caption{The same as Figure~\ref{fig:alphaeff} but for $\Mp=10^7\Msun$.}
\label{fig:alphaeff_M7}
\end{figure*}

\begin{figure*}
\centering
\includegraphics[width=\linewidth]{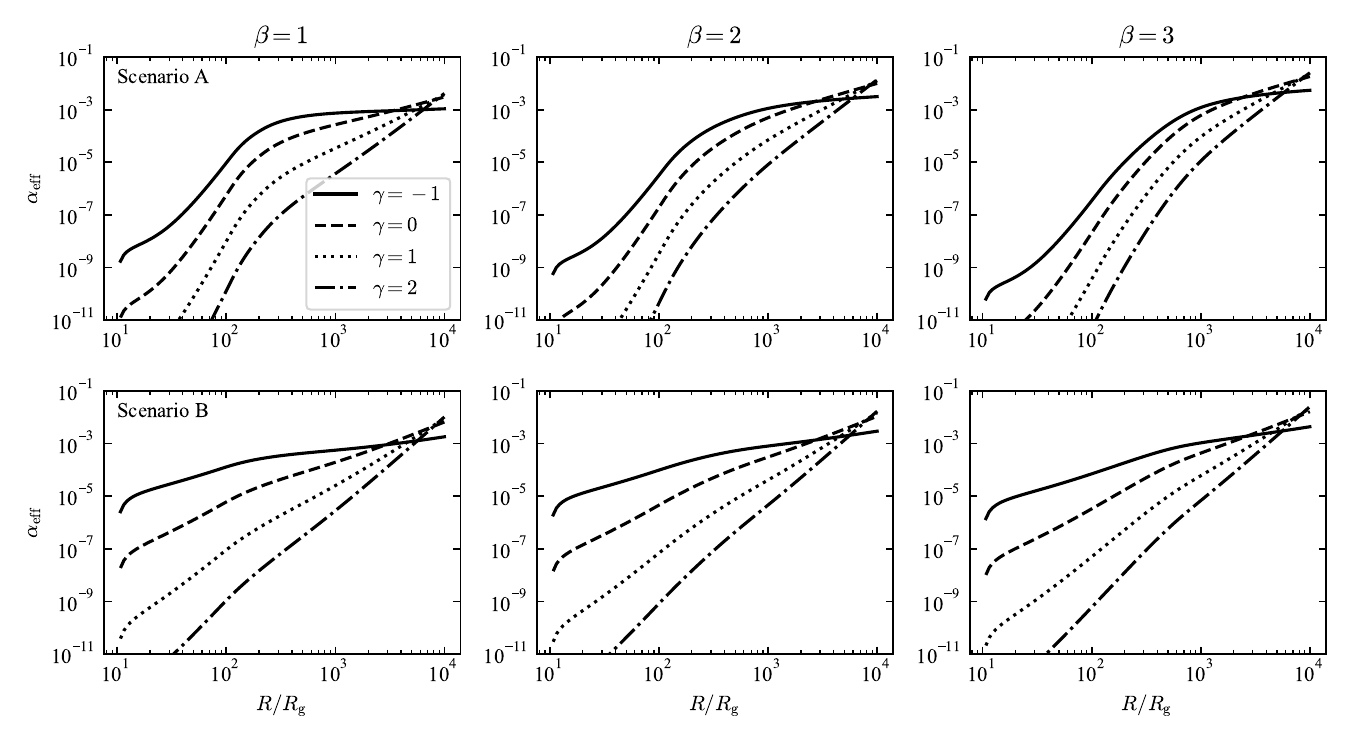}
\caption{The same as Figure~\ref{fig:alphaeff} but for $\dotMp=10$.}
\label{fig:alpha_dotM10}
\end{figure*}

\begin{figure*}
\centering
\includegraphics[width=\linewidth]{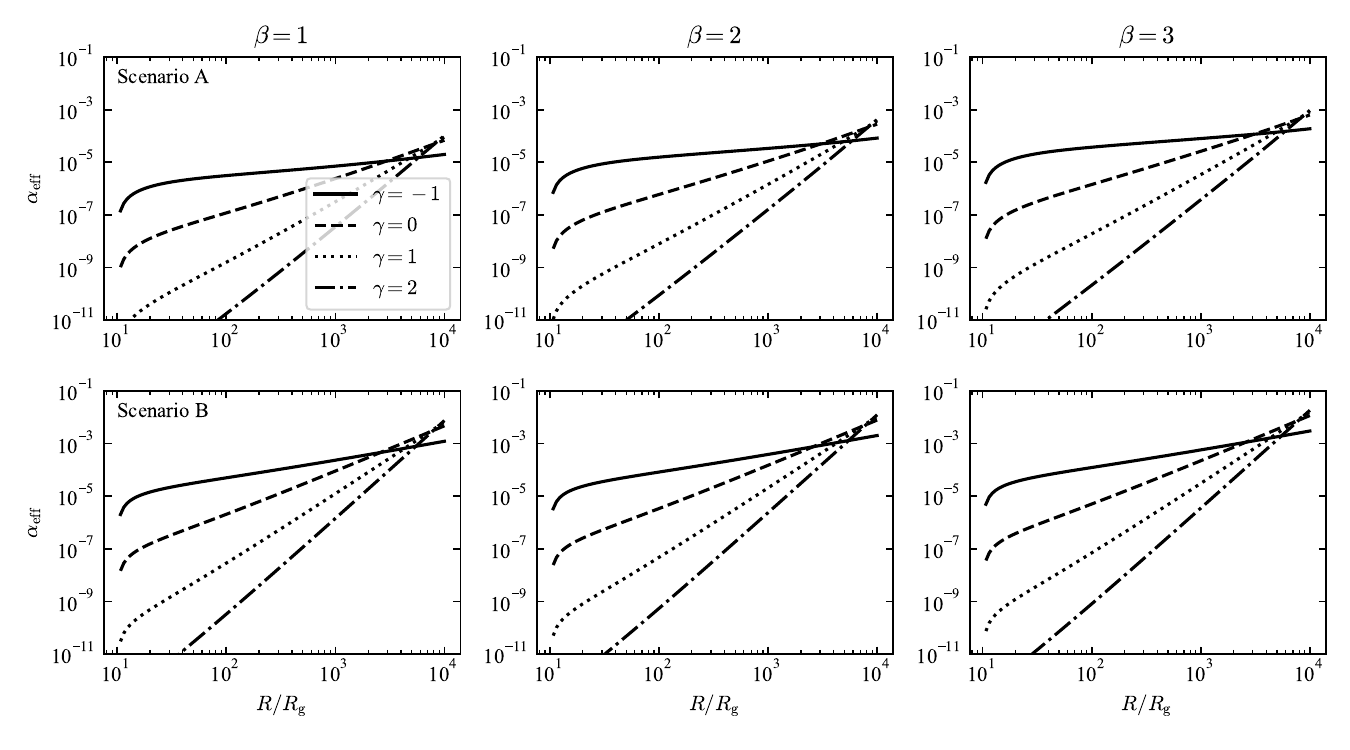}
\caption{The same as Figure~\ref{fig:alphaeff} but for $\dotMp=0.1$.}
\label{fig:alpha_dotM01}
\end{figure*}

We examine the influence of SMBH mass $\Mp$ and accretion rate $\dotMp$ on the effective viscosity $\alphaeff$ contributed by AMSs by plotting its radial profile under different disk parameters in Figures \ref{fig:alphaeff_M9}-\ref{fig:alpha_dotM01}.
The parameter sets are as follows:
Figure~\ref{fig:alphaeff_M9} adopts $\Mp = 10^9 \Msun$ and $\dotMp = 1$,
Figure~\ref{fig:alphaeff_M7} uses $\Mp = 10^7 \Msun$ and $\dotMp = 1$,
Figure~\ref{fig:alpha_dotM10} corresponds to $\Mp = 10^8 \Msun$ and $\dotMp = 10$,
and Figure~\ref{fig:alpha_dotM01} employs $\Mp = 10^8 \Msun$ and $\dotMp = 0.1$.
All figures show that $\alphaeff$ is significantly larger in the outer disk ($R/\Rg > 10^3$) than in the inner disk ($R/\Rg < 100$), supporting the conclusion that AMS serves as the source of viscosity in the outer disk for wide parameter space of $\Mp$ or $\dotMp$.

In Scenario A, for Figure~\ref{fig:alphaeff_M9} where $\Mp = 10^9 \Msun$, $\alphaeff$ is generally about two orders of magnitude greater than that in Figure~\ref{fig:alphaeff}.
Although gas density $\rho$ decreases, both the disk height $H$ and the Hill radius $\RHill$ increase, leading to more energy release (see Equation \ref{eq:dotMs}).
Conversely, in Figure~\ref{fig:alphaeff_M7} with $\Mp = 10^7 \Msun$, $\alphaeff$ is roughly two orders of magnitude smaller.
This supports the scaling $\alphaeff\propto\Mp^2$ as can be seen in Figure~\ref{fig:alphaAMS_M}.
In Figure~\ref{fig:alpha_dotM10} with $\dotMp = 10$, the higher $\rho$ and larger $H$ enhance the AMS outflow strength (Equation \ref{eq:dotMs}), raising $\alphaeff$ by about an order of magnitude in the outer disk compared to Figure~\ref{fig:alphaeff}.
In the inner region, however, $\alphaeff$ drops sharply due to decreasing $\rho$ and increasing sound speed $\cs$.
For Figure~\ref{fig:alpha_dotM01} with $\dotMp = 0.1$, lower $\rho$ and smaller $H$ weaken AMS outflows, reducing $\alphaeff$ by approximately one order of magnitude relative to Figure~\ref{fig:alphaeff}.
A rough proportionality $\alphaeff\propto\dotMp$ is found (also see Figure~\ref{fig:alphaAMS_dotM}).

In Scenario B, we find $\alphaeff\propto\Mp^{1.5}$ as can be seen in Figure~\ref{fig:alphaAMS_M},
as both $H$ and $\RHill$ increase with $\Mp$, enhancing the angular momentum transfer rate (see Equation \ref{eq:dotJB}).
In contrast, $\alphaeff\propto\dotMp^{0.1}$ shows almost no variation with the SMBH accretion rate (also see Figure~\ref{fig:alphaAMS_dotM}),
since $\RHill$ is independent of $\dotMp$ and $H\propto\dotMp^{3/20}$ is only weakly sensitive to $\dotMp$ in the outer region.

\clearpage

\clearpage
\bibliography{ref.bib}

@ARTICLE{Kitaki2021,
       author = {{Kitaki}, Takaaki and {Mineshige}, Shin and {Ohsuga}, Ken and {Kawashima}, Tomohisa},
        title = "{The origins and impact of outflow from super-Eddington flow}",
      journal = {\pasj},
         year = 2021,
        month = apr,
       volume = {73},
       number = {2},
        pages = {450-466},
          doi = {10.1093/pasj/psab011},
archivePrefix = {arXiv},
       eprint = {2101.11028},
 primaryClass = {astro-ph.HE},
       adsurl = {https://ui.adsabs.harvard.edu/abs/2021PASJ...73..450K}
}

@ARTICLE{Zhai2025,
       author = {{Zhai}, Shuo and {Wang}, Jian-Min and {Li}, Yan-Rong and {Guo}, Wei-Jian and {Zhao}, Gang},
        title = "{Chemical evolution of bulges of active galactic nuclei in the early Universe: roles of accreting stars}",
      journal = {arXiv e-prints},
         year = 2025,
        month = nov,
          eid = {arXiv:2511.16119},
        pages = {arXiv:2511.16119},
          doi = {10.48550/arXiv.2511.16119},
archivePrefix = {arXiv},
       eprint = {2511.16119},
 primaryClass = {astro-ph.GA},
       adsurl = {https://ui.adsabs.harvard.edu/abs/2025arXiv251116119Z}
}

@ARTICLE{Wang2025d,
       author = {{Wang}, Jian-Min and {Hu}, Chen and {Chen}, Yong-Jie and {Songsheng}, Yu-Yang and {Wang}, Yi-Lin and {Zhang}, Hao and {Du}, Pu and {Li}, Yan-Rong and {Luo}, Bin and {Brotherton}, Michael S. and {Bai}, Jin-Ming and {Guo}, Wei-Jian and {Yang}, Seng and {Yao}, Zhu-Heng and {Aceituno}, Jesus},
        title = "{Detection of unexpected leading delays in broad H{\ensuremath{\beta}} line reverberations in the quasar PHL 1092}",
      journal = {arXiv e-prints},
         year = 2025,
        month = nov,
          eid = {arXiv:2511.07716},
        pages = {arXiv:2511.07716},
          doi = {10.48550/arXiv.2511.07716},
archivePrefix = {arXiv},
       eprint = {2511.07716},
 primaryClass = {astro-ph.GA},
       adsurl = {https://ui.adsabs.harvard.edu/abs/2025arXiv251107716W}
}

@ARTICLE{Wang2025c,
       author = {{Wang}, Jian-Min and {Wang}, Yi-Lin and {Chen}, Yong-Jie and {Liu}, Jun-Rong and {Songsheng}, Yu-Yang and {Cheng}, Cheng and {Li}, Yan-Rong and {Du}, Pu and {Zhang}, Hao and {Zhao}, Yu},
        title = "{Little red dots as embryos of active galactic nuclei}",
      journal = {arXiv e-prints},
         year = 2025,
        month = nov,
          eid = {arXiv:2511.09278},
        pages = {arXiv:2511.09278},
          doi = {10.48550/arXiv.2511.09278},
archivePrefix = {arXiv},
       eprint = {2511.09278},
 primaryClass = {astro-ph.GA},
       adsurl = {https://ui.adsabs.harvard.edu/abs/2025arXiv251109278W}
}

@ARTICLE{Weaver1977,
       author = {{Weaver}, R. and {McCray}, R. and {Castor}, J. and {Shapiro}, P. and {Moore}, R.},
        title = "{Interstellar bubbles. II. Structure and evolution.}",
      journal = {\apj},
         year = 1977,
        month = dec,
       volume = {218},
        pages = {377-395},
          doi = {10.1086/155692},
       adsurl = {https://ui.adsabs.harvard.edu/abs/1977ApJ...218..377W}
}

@ARTICLE{Zhou2019,
       author = {{Zhou}, Yu and {Feng}, Hua and {Ho}, Luis C. and {Yao}, Yuhan},
        title = "{Evidence for Optically Thick, Eddington-limited Winds Driven by Supercritical Accretion}",
      journal = {\apj},
         year = 2019,
        month = jan,
       volume = {871},
       number = {1},
          eid = {115},
        pages = {115},
          doi = {10.3847/1538-4357/aaf724},
archivePrefix = {arXiv},
       eprint = {1812.02923},
 primaryClass = {astro-ph.HE},
       adsurl = {https://ui.adsabs.harvard.edu/abs/2019ApJ...871..115Z}
}

@ARTICLE{Yan2010,
       author = {{Yan}, Chang-Shuo and {Wang}, Jian-Min},
        title = "{Evolution of Gaseous Disk Viscosity Driven by Supernova Explosion. II. Structure and Emissions from Star-forming Galaxies at High Redshift}",
      journal = {\apj},
         year = 2010,
        month = dec,
       volume = {725},
       number = {2},
        pages = {2359-2380},
          doi = {10.1088/0004-637X/725/2/2359},
archivePrefix = {arXiv},
       eprint = {1010.4376},
 primaryClass = {astro-ph.CO},
       adsurl = {https://ui.adsabs.harvard.edu/abs/2010ApJ...725.2359Y}
}

@ARTICLE{Wang2009,
       author = {{Wang}, Jian-Min and {Yan}, Chang-Shuo and {Li}, Yan-Rong and {Chen}, Yan-Mei and {Xiang}, Fei and {Hu}, Chen and {Ge}, Jun-Qiang and {Zhang}, Shu},
        title = "{Evolution of Gaseous Disk Viscosity Driven by Supernova Explosions in Star-Forming Galaxies at High Redshift}",
      journal = {\apjl},
         year = 2009,
        month = aug,
       volume = {701},
       number = {1},
        pages = {L7-L11},
          doi = {10.1088/0004-637X/701/1/L7},
archivePrefix = {arXiv},
       eprint = {0907.4474},
 primaryClass = {astro-ph.CO},
       adsurl = {https://ui.adsabs.harvard.edu/abs/2009ApJ...701L...7W}
}

@ARTICLE{Wang2010,
       author = {{Wang}, Jian-Min and {Yan}, Chang-Shuo and {Gao}, Han-Qin and {Hu}, Chen and {Li}, Yan-Rong and {Zhang}, Shu},
        title = "{Accretion Disks in Active Galactic Nuclei: Gas Supply Driven by Star Formation}",
      journal = {\apjl},
         year = 2010,
        month = aug,
       volume = {719},
       number = {2},
        pages = {L148-L152},
          doi = {10.1088/2041-8205/719/2/L148},
archivePrefix = {arXiv},
       eprint = {1007.4060},
 primaryClass = {astro-ph.CO},
       adsurl = {https://ui.adsabs.harvard.edu/abs/2010ApJ...719L.148W}
}

@ARTICLE{Jiao2025,
       author = {{Jiao}, Cheng-Liang and {Zhu}, Liying and {Zhao}, Er-gang and {Zhang}, Jia},
        title = "{Accretion rates of stellar-mass compact objects embedded in AGN discs}",
      journal = {arXiv e-prints},
         year = 2025,
        month = oct,
          eid = {arXiv:2510.26111},
        pages = {arXiv:2510.26111},
archivePrefix = {arXiv},
       eprint = {2510.26111},
 primaryClass = {astro-ph.HE},
       adsurl = {https://ui.adsabs.harvard.edu/abs/2025arXiv251026111J}
}

@ARTICLE{Abramowicz1980,
       author = {{Abramowicz}, M.~A. and {Calvani}, M. and {Nobili}, L.},
        title = "{Thick accretion disks with super-Eddington luminosities}",
      journal = {\apj},
         year = 1980,
        month = dec,
       volume = {242},
        pages = {772-788},
          doi = {10.1086/158512},
       adsurl = {https://ui.adsabs.harvard.edu/abs/1980ApJ...242..772A}
}

@ARTICLE{Marshall2018,
       author = {{Marshall}, Megan D. and {Avara}, Mark J. and {McKinney}, Jonathan C.},
        title = "{Angular momentum transport in thin magnetically arrested discs}",
      journal = {\mnras},
         year = 2018,
        month = aug,
       volume = {478},
       number = {2},
        pages = {1837-1843},
          doi = {10.1093/mnras/sty1184},
archivePrefix = {arXiv},
       eprint = {1709.10113},
 primaryClass = {astro-ph.HE},
       adsurl = {https://ui.adsabs.harvard.edu/abs/2018MNRAS.478.1837M}
}

@ARTICLE{Kaisig1992,
       author = {{Kaisig}, M. and {Tajima}, T. and {Lovelace}, R.~V.~E.},
        title = "{Magnetic Interchange Instability of Accretion Disks}",
      journal = {\apj},
         year = 1992,
        month = feb,
       volume = {386},
        pages = {83},
          doi = {10.1086/170994},
       adsurl = {https://ui.adsabs.harvard.edu/abs/1992ApJ...386...83K}
}

@ARTICLE{Begelman2024,
       author = {{Begelman}, Mitchell C.},
        title = "{A simple model of globally magnetized accretion discs}",
      journal = {\mnras},
         year = 2024,
        month = nov,
       volume = {534},
       number = {4},
        pages = {3144-3154},
          doi = {10.1093/mnras/stae2305},
archivePrefix = {arXiv},
       eprint = {2402.15657},
 primaryClass = {astro-ph.HE},
       adsurl = {https://ui.adsabs.harvard.edu/abs/2024MNRAS.534.3144B}
}

@ARTICLE{Mishra2020,
       author = {{Mishra}, Bhupendra and {Begelman}, Mitchell C. and {Armitage}, Philip J. and {Simon}, Jacob B.},
        title = "{Strongly magnetized accretion discs: structure and accretion from global magnetohydrodynamic simulations}",
      journal = {\mnras},
         year = 2020,
        month = feb,
       volume = {492},
       number = {2},
        pages = {1855-1868},
          doi = {10.1093/mnras/stz3572},
archivePrefix = {arXiv},
       eprint = {1907.08995},
 primaryClass = {astro-ph.HE},
       adsurl = {https://ui.adsabs.harvard.edu/abs/2020MNRAS.492.1855M}
}

@ARTICLE{Goodman2003,
       author = {{Goodman}, Jeremy},
        title = "{Self-gravity and quasi-stellar object discs}",
      journal = {\mnras},
         year = 2003,
        month = mar,
       volume = {339},
       number = {4},
        pages = {937-948},
          doi = {10.1046/j.1365-8711.2003.06241.x},
archivePrefix = {arXiv},
       eprint = {astro-ph/0201001},
 primaryClass = {astro-ph},
       adsurl = {https://ui.adsabs.harvard.edu/abs/2003MNRAS.339..937G}
}

@ARTICLE{Ali2023,
       author = {{Ali-Dib}, Mohamad and {Lin}, Douglas N.~C.},
        title = "{The impermanent fate of massive stars in AGN discs}",
      journal = {\mnras},
         year = 2023,
        month = dec,
       volume = {526},
       number = {4},
        pages = {5824-5838},
          doi = {10.1093/mnras/stad2774},
archivePrefix = {arXiv},
       eprint = {2309.04392},
 primaryClass = {astro-ph.GA},
       adsurl = {https://ui.adsabs.harvard.edu/abs/2023MNRAS.526.5824A}
}

@ARTICLE{Luo2025,
       author = {{Luo}, Yang and {Wang}, Jian-Min},
        title = "{Numerical modelling the mass feeding rates on to accretion-modified stars embedded within AGN discs}",
      journal = {\mnras},
         year = 2025,
        month = jul,
       volume = {540},
       number = {3},
        pages = {2413-2421},
          doi = {10.1093/mnras/staf841},
archivePrefix = {arXiv},
       eprint = {2505.15048},
 primaryClass = {astro-ph.GA},
       adsurl = {https://ui.adsabs.harvard.edu/abs/2025MNRAS.540.2413L}
}

@ARTICLE{Xing2025,
       author = {{Xing}, Jing-Tong and {Liu}, Tong and {Huang}, Bao-Quan and {Sun}, Mouyuan},
        title = "{Variabilities Driven by Satellite Black Hole Migration in Active Galactic Nucleus Disks}",
      journal = {\apj},
         year = 2025,
        month = oct,
       volume = {991},
       number = {2},
          eid = {167},
        pages = {167},
          doi = {10.3847/1538-4357/adff71},
archivePrefix = {arXiv},
       eprint = {2501.10095},
 primaryClass = {astro-ph.HE},
       adsurl = {https://ui.adsabs.harvard.edu/abs/2025ApJ...991..167X}
}

@ARTICLE{Dutta2025,
       author = {{Dutta Roy}, Poulami and {Mahapatra}, Parthapratim and {Samajdar}, Anuradha and {Arun}, K.~G.},
        title = "{Identifying intermediate mass binary black hole mergers in AGN disks using LISA}",
      journal = {\prd},
         year = 2025,
        month = may,
       volume = {111},
       number = {10},
          eid = {104047},
        pages = {104047},
          doi = {10.1103/PhysRevD.111.104047},
archivePrefix = {arXiv},
       eprint = {2503.11721},
 primaryClass = {astro-ph.HE},
       adsurl = {https://ui.adsabs.harvard.edu/abs/2025PhRvD.111j4047D}
}

@ARTICLE{Gilbaum2025,
       author = {{Gilbaum}, Shmuel and {Grishin}, Evgeni and {Stone}, Nicholas C. and {Mandel}, Ilya},
        title = "{How to Escape from a Trap: Outcomes of Repeated Black Hole Mergers in Active Galactic Nuclei}",
      journal = {\apjl},
         year = 2025,
        month = mar,
       volume = {982},
       number = {1},
          eid = {L13},
        pages = {L13},
          doi = {10.3847/2041-8213/adb7dc},
archivePrefix = {arXiv},
       eprint = {2410.19904},
 primaryClass = {astro-ph.HE},
       adsurl = {https://ui.adsabs.harvard.edu/abs/2025ApJ...982L..13G}
}

@ARTICLE{Fabj2024,
       author = {{Fabj}, Gaia and {Samsing}, Johan},
        title = "{Eccentric mergers in AGN discs: influence of the supermassive black hole on three-body interactions}",
      journal = {\mnras},
         year = 2024,
        month = dec,
       volume = {535},
       number = {4},
        pages = {3630-3645},
          doi = {10.1093/mnras/stae2499},
archivePrefix = {arXiv},
       eprint = {2402.16948},
 primaryClass = {astro-ph.GA},
       adsurl = {https://ui.adsabs.harvard.edu/abs/2024MNRAS.535.3630F}
}

@ARTICLE{Samsing2025,
       author = {{Samsing}, Johan and {Hendriks}, Kai and {Zwick}, Lorenz and {D'Orazio}, Daniel J. and {Liu}, Bin},
        title = "{Gravitational-wave Phase Shifts in Eccentric Black Hole Mergers as a Probe of Dynamical Formation Environments}",
      journal = {\apj},
         year = 2025,
        month = sep,
       volume = {990},
       number = {2},
          eid = {211},
        pages = {211},
          doi = {10.3847/1538-4357/ad9f3d},
archivePrefix = {arXiv},
       eprint = {2403.05625},
 primaryClass = {astro-ph.HE},
       adsurl = {https://ui.adsabs.harvard.edu/abs/2025ApJ...990..211S}
}

@ARTICLE{Zhang2025Lizhong,
       author = {{Zhang}, Lizhong and {Stone}, James M. and {White}, Christopher J. and {Davis}, Shane W. and {Jiang}, Yan-Fei and {Mullen}, Patrick D.},
        title = "{Radiation GRMHD Models of Accretion onto Stellar-Mass Black Holes: II. Super-Eddington Accretion}",
      journal = {arXiv e-prints},
         year = 2025,
        month = sep,
          eid = {arXiv:2509.10638},
        pages = {arXiv:2509.10638},
          doi = {10.48550/arXiv.2509.10638},
archivePrefix = {arXiv},
       eprint = {2509.10638},
 primaryClass = {astro-ph.HE},
       adsurl = {https://ui.adsabs.harvard.edu/abs/2025arXiv250910638Z}
}

@ARTICLE{Xu2025,
       author = {{Xu}, Zheng-Hao},
        title = "{The Fate of Stars Embedded in AGN Disks is Determined by an Internal Mixing Threshold}",
      journal = {Research in Astronomy and Astrophysics},
         year = 2025,
        month = nov,
       volume = {25},
       number = {11},
          eid = {115013},
        pages = {115013},
          doi = {10.1088/1674-4527/adfeb9},
       adsurl = {https://ui.adsabs.harvard.edu/abs/2025RAA....25k5013X}
}

@ARTICLE{Li2023,
       author = {{Li}, Fu-Lin and {Liu}, Yu and {Fan}, Xiao and {Hu}, Mao-Kai and {Yang}, Xuan and {Geng}, Jin-Jun and {Wu}, Xue-Feng},
        title = "{Core-collapse Supernova Explosions in Active Galactic Nucleus Accretion Disks}",
      journal = {\apj},
         year = 2023,
        month = jun,
       volume = {950},
       number = {2},
          eid = {161},
        pages = {161},
          doi = {10.3847/1538-4357/acd2d1},
archivePrefix = {arXiv},
       eprint = {2305.04010},
 primaryClass = {astro-ph.HE},
       adsurl = {https://ui.adsabs.harvard.edu/abs/2023ApJ...950..161L}
}

@ARTICLE{She2025,
       author = {{She}, Jiao-Zhen and {Liu}, Tong and {Huang}, Bao-Quan and {Wei}, Yun-Feng and {Li}, Fu-Lin and {Geng}, Jin-Jun and {Wu}, Xue-Feng},
        title = "{Anisotropy of Core-collapse Supernovae Effected by the Disks of Active Galactic Nuclei}",
      journal = {\apj},
         year = 2025,
        month = sep,
       volume = {990},
       number = {1},
          eid = {17},
        pages = {17},
          doi = {10.3847/1538-4357/adefe6},
archivePrefix = {arXiv},
       eprint = {2507.10040},
 primaryClass = {astro-ph.HE},
       adsurl = {https://ui.adsabs.harvard.edu/abs/2025ApJ...990...17S}
}

@ARTICLE{Postiglione2025,
       author = {{Postiglione}, Jake and {Ford}, K.~E. Saavik and {Best}, Henry and {McKernan}, Barry and {O'Dowd}, Matthew},
        title = "{Evolution of LISA Observables for Binary Black Holes Lensed by a Supermassive Black Hole}",
      journal = {\apj},
         year = 2025,
        month = oct,
       volume = {991},
       number = {2},
          eid = {161},
        pages = {161},
          doi = {10.3847/1538-4357/adfa0c},
archivePrefix = {arXiv},
       eprint = {2502.10591},
 primaryClass = {astro-ph.HE},
       adsurl = {https://ui.adsabs.harvard.edu/abs/2025ApJ...991..161P}
}

@ARTICLE{McKernan2025,
       author = {{McKernan}, Barry and {Ford}, K.~E. Saavik and {Cook}, Harrison E. and {Delfavero}, Vera and {McPike}, Emily and {Nathaniel}, Kaila and {Postiglione}, Jake and {Ray}, Shawn and {O'Shaughnessy}, Richard},
        title = "{McFACTS I: Testing the LVK AGN Channel with Monte Carlo for AGN Channel Testing and Simulation (McFACTS)}",
      journal = {\apj},
         year = 2025,
        month = sep,
       volume = {990},
       number = {2},
          eid = {217},
        pages = {217},
          doi = {10.3847/1538-4357/adf114},
archivePrefix = {arXiv},
       eprint = {2410.16515},
 primaryClass = {astro-ph.HE},
       adsurl = {https://ui.adsabs.harvard.edu/abs/2025ApJ...990..217M}
}

@ARTICLE{Whitehead2025b,
       author = {{Whitehead}, Henry and {Rowan}, Connar and {Kocsis}, Bence},
        title = "{Hydrodynamic simulations of black hole evolution in AGN discs II: inclination damping for partially embedded satellites}",
      journal = {\mnras},
         year = 2025,
        month = oct,
          doi = {10.1093/mnras/staf1686},
archivePrefix = {arXiv},
       eprint = {2505.23899},
 primaryClass = {astro-ph.HE},
       adsurl = {https://ui.adsabs.harvard.edu/abs/2025MNRAS.tmp.1587W}
}

@ARTICLE{Rowan2025,
       author = {{Rowan}, Connar and {Whitehead}, Henry and {Fabj}, Gaia and {Kirkeberg}, Philip and {Pessah}, Martin E. and {Kocsis}, Bence},
        title = "{Hydrodynamic simulations of black hole evolution in AGN discs {\textendash} I. Orbital alignment of highly inclined satellites}",
      journal = {\mnras},
         year = 2025,
        month = oct,
       volume = {543},
       number = {1},
        pages = {132-145},
          doi = {10.1093/mnras/staf1449},
archivePrefix = {arXiv},
       eprint = {2505.23739},
 primaryClass = {astro-ph.HE},
       adsurl = {https://ui.adsabs.harvard.edu/abs/2025MNRAS.543..132R}
}

@ARTICLE{Trani2025,
       author = {{Trani}, Alessandro Alberto and {Di Cintio}, Pierfrancesco},
        title = "{Turbulent drag on stellar mass black holes embedded in AGN discs}",
      journal = {arXiv e-prints},
         year = 2025,
        month = jun,
          eid = {arXiv:2506.02173},
        pages = {arXiv:2506.02173},
          doi = {10.48550/arXiv.2506.02173},
archivePrefix = {arXiv},
       eprint = {2506.02173},
 primaryClass = {astro-ph.HE},
       adsurl = {https://ui.adsabs.harvard.edu/abs/2025arXiv250602173T}
}

@ARTICLE{Whitehead2025,
       author = {{Whitehead}, Henry and {Rowan}, Connar and {Kocsis}, Bence},
        title = "{3D adiabatic simulations of binary black hole formation in AGN discs}",
      journal = {\mnras},
         year = 2025,
        month = sep,
       volume = {542},
       number = {2},
        pages = {1033-1055},
          doi = {10.1093/mnras/staf1271},
archivePrefix = {arXiv},
       eprint = {2502.14959},
 primaryClass = {astro-ph.HE},
       adsurl = {https://ui.adsabs.harvard.edu/abs/2025MNRAS.542.1033W}
}

@ARTICLE{Fabj2025b,
       author = {{Fabj}, Gaia and {Tiede}, Christopher and {Rowan}, Connar and {Pessah}, Martin and {Samsing}, Johan},
        title = "{Spin-Orbit Misalignments of Eccentric Black Hole Mergers in AGN Disks}",
      journal = {arXiv e-prints},
         year = 2025,
        month = oct,
          eid = {arXiv:2510.07952},
        pages = {arXiv:2510.07952},
          doi = {10.48550/arXiv.2510.07952},
archivePrefix = {arXiv},
       eprint = {2510.07952},
 primaryClass = {astro-ph.HE},
       adsurl = {https://ui.adsabs.harvard.edu/abs/2025arXiv251007952F}
}

@ARTICLE{Zhou2024,
       author = {{Zhou}, Shuying and {Sun}, Mouyuan and {Liu}, Tong and {Wang}, Jian-Min and {Wang}, Jun-Xian and {Xue}, Yongquan},
        title = "{Stellar Black Holes Can ``Stretch'' Supermassive Black Hole Accretion Disks}",
      journal = {\apjl},
         year = 2024,
        month = may,
       volume = {966},
       number = {1},
          eid = {L9},
        pages = {L9},
          doi = {10.3847/2041-8213/ad3c3f},
archivePrefix = {arXiv},
       eprint = {2404.07407},
 primaryClass = {astro-ph.HE},
       adsurl = {https://ui.adsabs.harvard.edu/abs/2024ApJ...966L...9Z}
}

@ARTICLE{Kishor2025,
       author = {{Kishor Joshi}, Raj and {Bhake}, Aryan and {Banerjee}, Biswajit and {Vaidya}, Bhargav and {Ruiz}, Milton and {Tsokaros}, Antonios and {Mignone}, Andrea and {Branchesi}, Marica and {Shukla}, Amit and {{\v{C}}emelji{\'c}}, Miljenko},
        title = "{Binary black holes in magnetized AGN disks}",
      journal = {arXiv e-prints},
         year = 2025,
        month = sep,
          eid = {arXiv:2509.16796},
        pages = {arXiv:2509.16796},
          doi = {10.48550/arXiv.2509.16796},
archivePrefix = {arXiv},
       eprint = {2509.16796},
 primaryClass = {astro-ph.HE},
       adsurl = {https://ui.adsabs.harvard.edu/abs/2025arXiv250916796K}
}

@ARTICLE{Abhishek2025,
       author = {{Abhishek Hegade K.}, R. and {Gammie}, Charles F. and {Yunes}, Nicol{\'a}s},
        title = "{A relativistic treatment of accretion disk torques on extreme mass ratio inspirals around spinning black holes}",
      journal = {arXiv e-prints},
         year = 2025,
        month = oct,
          eid = {arXiv:2510.03564},
        pages = {arXiv:2510.03564},
          doi = {10.48550/arXiv.2510.03564},
archivePrefix = {arXiv},
       eprint = {2510.03564},
 primaryClass = {gr-qc},
       adsurl = {https://ui.adsabs.harvard.edu/abs/2025arXiv251003564A}
}

@ARTICLE{Duque2025,
       author = {{Duque}, Francisco and {Sberna}, Laura and {Spiers}, Andrew and {Vicente}, Rodrigo},
        title = "{Extreme-mass-ratio inspirals in relativistic accretion discs}",
      journal = {arXiv e-prints},
         year = 2025,
        month = oct,
          eid = {arXiv:2510.02433},
        pages = {arXiv:2510.02433},
          doi = {10.48550/arXiv.2510.02433},
archivePrefix = {arXiv},
       eprint = {2510.02433},
 primaryClass = {gr-qc},
       adsurl = {https://ui.adsabs.harvard.edu/abs/2025arXiv251002433D}
}

@ARTICLE{Sun2025,
       author = {{Sun}, Houyi and {Li}, Ya-Ping and {Pan}, Zhen and {Yang}, Huan},
        title = "{Probing Formation Channels of Extreme Mass-Ratio Inspirals}",
      journal = {arXiv e-prints},
         year = 2025,
        month = aug,
          eid = {arXiv:2509.00469},
        pages = {arXiv:2509.00469},
          doi = {10.48550/arXiv.2509.00469},
archivePrefix = {arXiv},
       eprint = {2509.00469},
 primaryClass = {gr-qc},
       adsurl = {https://ui.adsabs.harvard.edu/abs/2025arXiv250900469S}
}

@ARTICLE{Wang2025a,
       author = {{Wang}, Mengye and {Ma}, Yiqiu and {Li}, Hui and {Wu}, Qingwen and {Li}, Ya-Ping and {Lei}, Xiangli and {Wu}, Jiancheng},
        title = "{Simulation of Binary-single Interactions in AGN Disk. I. Gas-enhanced Binary Orbital Hardening}",
      journal = {\apj},
         year = 2025,
        month = apr,
       volume = {983},
       number = {2},
          eid = {114},
        pages = {114},
          doi = {10.3847/1538-4357/adbf8e},
archivePrefix = {arXiv},
       eprint = {2501.10703},
 primaryClass = {astro-ph.HE},
       adsurl = {https://ui.adsabs.harvard.edu/abs/2025ApJ...983..114W}
}

@ARTICLE{Wang2025b,
       author = {{Wang}, Mengye and {Wu}, Qingwen and {Ma}, Yiqiu},
        title = "{Simulation of Binary-Single Interactions in AGN Disks II: Merger Probability of Binary Black Holes during Chaotic Triple Process}",
      journal = {arXiv e-prints},
         year = 2025,
        month = jul,
          eid = {arXiv:2507.07715},
        pages = {arXiv:2507.07715},
          doi = {10.48550/arXiv.2507.07715},
archivePrefix = {arXiv},
       eprint = {2507.07715},
 primaryClass = {astro-ph.HE},
       adsurl = {https://ui.adsabs.harvard.edu/abs/2025arXiv250707715W}
}

@ARTICLE{Dittmann2025b,
       author = {{Dittmann}, Alexander J. and {Dempsey}, Adam M. and {Li}, Hui},
        title = "{The Multiple Paths to Merger of Unequal-mass Black Hole Binaries in the Disks of Active Galactic Nuclei}",
      journal = {\apj},
         year = 2025,
        month = sep,
       volume = {990},
       number = {2},
          eid = {137},
        pages = {137},
          doi = {10.3847/1538-4357/adea72},
archivePrefix = {arXiv},
       eprint = {2505.05555},
 primaryClass = {astro-ph.HE},
       adsurl = {https://ui.adsabs.harvard.edu/abs/2025ApJ...990..137D}
}

@ARTICLE{Dittmann2025,
       author = {{Dittmann}, Alexander J. and {Cantiello}, Matteo},
        title = "{A Semi-analytical Model for Stellar Evolution in AGN Disks}",
      journal = {\apj},
         year = 2025,
        month = feb,
       volume = {979},
       number = {2},
          eid = {245},
        pages = {245},
          doi = {10.3847/1538-4357/ad9e92},
archivePrefix = {arXiv},
       eprint = {2409.02981},
 primaryClass = {astro-ph.GA},
       adsurl = {https://ui.adsabs.harvard.edu/abs/2025ApJ...979..245D}
}

@ARTICLE{Liu2025a,
       author = {{Liu}, Jun-Rong and {Wang}, Jian-Min and {Fermi-LAT Collaboration} and {Abdollahi}, S. and {Ajello}, M. and {Batista}, R. Alves and {Baldini}, L. and {Bartolini}, C. and {Bastieri}, D. and {Gonzalez}, J. Becerra and {Bellazzini}, R. and {Berenji}, B. and {Bissaldi}, E. and {Blandford}, R.~D. and {Bonino}, R. and {Bruel}, P. and {Buson}, S. and {Cameron}, R.~A. and {Caraveo}, P.~A. and {Cavazzuti}, E. and {Chiaro}, G. and {Cibrario}, N. and {Ciprini}, S. and {Cristarella Orestano}, P. and {Cutini}, S. and {D'Ammando}, F. and {Di Lalla}, N. and {Dinesh}, A. and {Di Venere}, L. and {Dom{\'\i}nguez}, A. and {Fegan}, S.~J. and {Fiori}, A. and {Franckowiak}, A. and {Fukazawa}, Y. and {Funk}, S. and {Fusco}, P. and {Gargano}, F. and {Gasbarra}, C. and {Gasparrini}, D. and {Germani}, S. and {Giglietto}, N. and {Giliberti}, M. and {Giordano}, F. and {Giroletti}, M. and {Green}, D. and {Grenier}, I.~A. and {Guiriec}, S. and {Hashizume}, M. and {Hays}, E. and {Hewitt}, J.~W. and {Horan}, D. and {Hou}, Xian and {Karwin}, C. and {Kayanoki}, T. and {Kuss}, M. and {Laviron}, A. and {Lemoine-Goumard}, M. and {Li}, Jian and {Liodakis}, I. and {Longo}, F. and {Loparco}, F. and {Lorusso}, L. and {Lubrano}, P. and {Maldera}, S. and {Marcotulli}, L. and {Mart{\'\i}-Devesa}, G. and {Mazziotta}, M.~N. and {Mereu}, I. and {Michelson}, P.~F. and {Mirabal}, N. and {Mitthumsiri}, W. and {Mizuno}, T. and {Monzani}, M.~E. and {Morishita}, T. and {Morselli}, A. and {Moskalenko}, I.~V. and {Negro}, M. and {Niwa}, R. and {Omodei}, N. and {Orienti}, M. and {Orlando}, E. and {Ormes}, J.~F. and {Paneque}, D. and {Panzarini}, G. and {Persic}, M. and {Pesce-Rollins}, M. and {Pillera}, R. and {Porter}, T.~A. and {Principe}, G. and {Rain{\`o}}, S. and {Rando}, R. and {Rani}, B. and {Razzano}, M. and {Reimer}, A. and {Reimer}, O. and {S{\'a}nchez-Conde}, M. and {Saz Parkinson}, P.~M. and {Serini}, D. and {Sgr{\`o}}, C. and {Siskind}, E.~J. and {Spandre}, G. and {Spinelli}, P. and {Suson}, D.~J. and {Tajima}, H. and {Thayer}, J.~B. and {Torres}, D.~F. and {Zhao}, Zi-Hao},
        title = "{Fermi detection of gamma-ray emission from the hot coronae of radio-quiet active galactic nuclei}",
      journal = {Nature Astronomy},
         year = 2025,
        month = jul,
       volume = {9},
        pages = {1086-1097},
          doi = {10.1038/s41550-025-02538-2},
archivePrefix = {arXiv},
       eprint = {2502.19189},
 primaryClass = {astro-ph.HE},
       adsurl = {https://ui.adsabs.harvard.edu/abs/2025NatAs...9.1086L}
}

@ARTICLE{Liu2025b,
       author = {{Liu}, Jun-Rong and {Feng}, Hua and {Wang}, Jian-Min},
        title = "{Fermi detection of $γ$-rays from the radio-quiet Seyfert galaxy NGC 3281}",
      journal = {arXiv e-prints},
         year = 2025,
        month = sep,
          eid = {arXiv:2509.21958},
        pages = {arXiv:2509.21958},
          doi = {10.48550/arXiv.2509.21958},
archivePrefix = {arXiv},
       eprint = {2509.21958},
 primaryClass = {astro-ph.HE},
       adsurl = {https://ui.adsabs.harvard.edu/abs/2025arXiv250921958L}
}

@ARTICLE{Epstein2025,
       author = {{Epstein-Martin}, Marguerite and {Tagawa}, Hiromichi and {Haiman}, Zolt{\'a}n and {Perna}, Rosalba},
        title = "{Time-dependent models of AGN discs with radiation from embedded stellar-mass black holes}",
      journal = {\mnras},
         year = 2025,
        month = mar,
       volume = {537},
       number = {4},
        pages = {3396-3420},
          doi = {10.1093/mnras/staf237},
archivePrefix = {arXiv},
       eprint = {2405.09380},
 primaryClass = {astro-ph.HE},
       adsurl = {https://ui.adsabs.harvard.edu/abs/2025MNRAS.537.3396E}
}

@ARTICLE{Abbott2020b,
       author = {{Abbott}, R. and {Abbott}, T.~D. and {Abraham}, S. and {Acernese}, F. and {Ackley}, K. and {Adams}, C. and {Adhikari}, R.~X. and {Adya}, V.~B. and {Affeldt}, C. and {Agathos}, M. and {Agatsuma}, K. and {Aggarwal}, N. and {Aguiar}, O.~D. and {Aich}, A. and {Aiello}, L. and {Ain}, A. and {Ajith}, P. and {Akcay}, S. and {Allen}, G. and {Allocca}, A. and {Altin}, P.~A. and {Amato}, A. and {Anand}, S. and {Ananyeva}, A. and {Anderson}, S.~B. and {Anderson}, W.~G. and {Angelova}, S.~V. and {Ansoldi}, S. and {Antier}, S. and {Appert}, S. and {Arai}, K. and {Araya}, M.~C. and {Areeda}, J.~S. and {Ar{\`e}ne}, M. and {Arnaud}, N. and {Aronson}, S.~M. and {Arun}, K.~G. and {Asali}, Y. and {Ascenzi}, S. and {Ashton}, G. and {Aston}, S.~M. and {Astone}, P. and {Aubin}, F. and {Aufmuth}, P. and {AultONeal}, K. and {Austin}, C. and {Avendano}, V. and {Babak}, S. and {Bacon}, P. and {Badaracco}, F. and {Bader}, M.~K.~M. and {Bae}, S. and {Baer}, A.~M. and {Baird}, J. and {Baldaccini}, F. and {Ballardin}, G. and {Ballmer}, S.~W. and {Bals}, A. and {Balsamo}, A. and {Baltus}, G. and {Banagiri}, S. and {Bankar}, D. and {Bankar}, R.~S. and {Barayoga}, J.~C. and {Barbieri}, C. and {Barish}, B.~C. and {Barker}, D. and {Barkett}, K. and {Barneo}, P. and {Barone}, F. and {Barr}, B. and {Barsotti}, L. and {Barsuglia}, M. and {Barta}, D. and {Bartlett}, J. and {Bartos}, I. and {Bassiri}, R. and {Basti}, A. and {Bawaj}, M. and {Bayley}, J.~C. and {Bazzan}, M. and {B{\'e}csy}, B. and {Bejger}, M. and {Belahcene}, I. and {Bell}, A.~S. and {Beniwal}, D. and {Benjamin}, M.~G. and {Benkel}, R. and {Bentley}, J.~D. and {Bergamin}, F. and {Berger}, B.~K. and {Bergmann}, G. and {Bernuzzi}, S. and {Berry}, C.~P.~L. and {Bersanetti}, D. and {Bertolini}, A. and {Betzwieser}, J. and {Bhandare}, R. and {Bhandari}, A.~V. and {Bidler}, J. and {Biggs}, E. and {Bilenko}, I.~A. and {Billingsley}, G. and {Birney}, R. and {Birnholtz}, O. and {Biscans}, S. and {Bischi}, M. and {Biscoveanu}, S. and {Bisht}, A. and {Bissenbayeva}, G. and {Bitossi}, M. and {Bizouard}, M.~A. and {Blackburn}, J.~K. and {Blackman}, J. and {Blair}, C.~D. and {Blair}, D.~G. and {Blair}, R.~M. and {Bobba}, F. and {Bode}, N. and {Boer}, M. and {Boetzel}, Y. and {Bogaert}, G. and {Bondu}, F. and {Bonilla}, E. and {Bonnand}, R. and {Booker}, P. and {Boom}, B.~A. and {Bork}, R. and {Boschi}, V. and {Bose}, S. and {Bossilkov}, V. and {Bosveld}, J. and {Bouffanais}, Y. and {Bozzi}, A. and {Bradaschia}, C. and {Brady}, P.~R. and {Bramley}, A. and {Branchesi}, M. and {Brau}, J.~E. and {Breschi}, M. and {Briant}, T. and {Briggs}, J.~H. and {Brighenti}, F. and {Brillet}, A. and {Brinkmann}, M. and {Brito}, R. and {Brockill}, P. and {Brooks}, A.~F. and {Brooks}, J. and {Brown}, D.~D. and {Brunett}, S. and {Bruno}, G. and {Bruntz}, R. and {Buikema}, A. and {Bulik}, T. and {Bulten}, H.~J. and {Buonanno}, A. and {Buskulic}, D. and {Byer}, R.~L. and {Cabero}, M. and {Cadonati}, L. and {Cagnoli}, G. and {Cahillane}, C. and {Bustillo}, J. Calder{\'o}n and {Callaghan}, J.~D. and {Callister}, T.~A. and {Calloni}, E. and {Camp}, J.~B. and {Canepa}, M. and {Cannon}, K.~C. and {Cao}, H. and {Cao}, J. and {Carapella}, G. and {Carbognani}, F. and {Caride}, S. and {Carney}, M.~F. and {Carullo}, G. and {Diaz}, J. Casanueva and {Casentini}, C. and {Casta{\~n}eda}, J. and {Caudill}, S. and {Cavagli{\`a}}, M. and {Cavalier}, F. and {Cavalieri}, R. and {Cella}, G. and {Cerd{\'a}-Dur{\'a}n}, P. and {Cesarini}, E. and {Chaibi}, O. and {Chakravarti}, K. and {Chan}, C. and {Chan}, M. and {Chao}, S. and {Charlton}, P. and {Chase}, E.~A. and {Chassande-Mottin}, E. and {Chatterjee}, D. and {Chaturvedi}, M. and {Chatziioannou}, K. and {Chen}, H.~Y. and {Chen}, X.},
        title = "{GW190814: Gravitational Waves from the Coalescence of a 23 Solar Mass Black Hole with a 2.6 Solar Mass Compact Object}",
      journal = {\apjl},
         year = 2020,
        month = jun,
       volume = {896},
       number = {2},
          eid = {L44},
        pages = {L44},
          doi = {10.3847/2041-8213/ab960f},
archivePrefix = {arXiv},
       eprint = {2006.12611},
 primaryClass = {astro-ph.HE},
       adsurl = {https://ui.adsabs.harvard.edu/abs/2020ApJ...896L..44A}
}

@ARTICLE{Delfavero2025,
       author = {{Delfavero}, V. and {Ray}, S. and {Cook}, H.~E. and {Nathaniel}, K. and {McKernan}, B. and {Ford}, K.~E.~S. and {Postiglione}, J. and {McPike}, E. and {O'Shaughnessy}, R.},
        title = "{Prospects for the formation of GW231123 from the AGN channel}",
      journal = {arXiv e-prints},
         year = 2025,
        month = aug,
          eid = {arXiv:2508.13412},
        pages = {arXiv:2508.13412},
          doi = {10.48550/arXiv.2508.13412},
archivePrefix = {arXiv},
       eprint = {2508.13412},
 primaryClass = {gr-qc},
       adsurl = {https://ui.adsabs.harvard.edu/abs/2025arXiv250813412D}
}

@ARTICLE{Ford2025,
       author = {{Ford}, K.~E. Saavik and {McKernan}, Barry},
        title = "{Using gravitational waves and multi-messenger Astronomy to reverse-engineer the properties of galactic nuclei}",
      journal = {arXiv e-prints},
         year = 2025,
        month = jun,
          eid = {arXiv:2506.08801},
        pages = {arXiv:2506.08801},
          doi = {10.48550/arXiv.2506.08801},
archivePrefix = {arXiv},
       eprint = {2506.08801},
 primaryClass = {astro-ph.HE},
       adsurl = {https://ui.adsabs.harvard.edu/abs/2025arXiv250608801F}
}

@ARTICLE{Vaccaro2025b,
       author = {{Vaccaro}, Maria Paola},
        title = "{Hierarchical Black Hole Mergers in AGN Disks: Tracing Massive Black Hole Growth Across Cosmic Time}",
      journal = {arXiv e-prints},
         year = 2025,
        month = aug,
          eid = {arXiv:2508.15337},
        pages = {arXiv:2508.15337},
          doi = {10.48550/arXiv.2508.15337},
archivePrefix = {arXiv},
       eprint = {2508.15337},
 primaryClass = {astro-ph.HE},
       adsurl = {https://ui.adsabs.harvard.edu/abs/2025arXiv250815337V}
}

@ARTICLE{Vaccaro2025,
       author = {{Vaccaro}, Maria Paola and {Seif}, Yannick and {Mapelli}, Michela},
        title = "{The role of migration traps in the formation of binary black holes in AGN disks}",
      journal = {arXiv e-prints},
         year = 2025,
        month = aug,
          eid = {arXiv:2508.03637},
        pages = {arXiv:2508.03637},
          doi = {10.48550/arXiv.2508.03637},
archivePrefix = {arXiv},
       eprint = {2508.03637},
 primaryClass = {astro-ph.HE},
       adsurl = {https://ui.adsabs.harvard.edu/abs/2025arXiv250803637V}
}

@ARTICLE{Li2025,
       author = {{Li}, Yin-Jie and {Wang}, Yuan-Zhu and {Tang}, Shao-Peng and {Fan}, Yi-Zhong},
        title = "{Aligned Hierarchical Black Hole Mergers in AGN disks revealed by GWTC-4}",
      journal = {arXiv e-prints},
         year = 2025,
        month = sep,
          eid = {arXiv:2509.23897},
        pages = {arXiv:2509.23897},
          doi = {10.48550/arXiv.2509.23897},
archivePrefix = {arXiv},
       eprint = {2509.23897},
 primaryClass = {astro-ph.HE},
       adsurl = {https://ui.adsabs.harvard.edu/abs/2025arXiv250923897L}
}

@ARTICLE{Xue2025,
       author = {{Xue}, LingQin and {Tagawa}, Hiromichi and {Haiman}, Zolt{\'a}n and {Bartos}, Imre},
        title = "{What determines the maximum mass of AGN-assisted black hole mergers?}",
      journal = {\prd},
         year = 2025,
        month = sep,
       volume = {112},
       number = {6},
          eid = {063034},
        pages = {063034},
          doi = {10.1103/5m1n-qh9v},
archivePrefix = {arXiv},
       eprint = {2504.19570},
 primaryClass = {astro-ph.HE},
       adsurl = {https://ui.adsabs.harvard.edu/abs/2025PhRvD.112f3034X}
}

@ARTICLE{Gilbaum2022,
       author = {{Gilbaum}, Shmuel and {Stone}, Nicholas C.},
        title = "{Feedback-dominated Accretion Flows}",
      journal = {\apj},
         year = 2022,
        month = apr,
       volume = {928},
       number = {2},
          eid = {191},
        pages = {191},
          doi = {10.3847/1538-4357/ac4ded},
archivePrefix = {arXiv},
       eprint = {2107.07519},
 primaryClass = {astro-ph.HE},
       adsurl = {https://ui.adsabs.harvard.edu/abs/2022ApJ...928..191G}
}

@ARTICLE{Babak2017,
       author = {{Babak}, Stanislav and {Gair}, Jonathan and {Sesana}, Alberto and {Barausse}, Enrico and {Sopuerta}, Carlos F. and {Berry}, Christopher P.~L. and {Berti}, Emanuele and {Amaro-Seoane}, Pau and {Petiteau}, Antoine and {Klein}, Antoine},
        title = "{Science with the space-based interferometer LISA. V. Extreme mass-ratio inspirals}",
      journal = {\prd},
         year = 2017,
        month = may,
       volume = {95},
       number = {10},
          eid = {103012},
        pages = {103012},
          doi = {10.1103/PhysRevD.95.103012},
archivePrefix = {arXiv},
       eprint = {1703.09722},
 primaryClass = {gr-qc},
       adsurl = {https://ui.adsabs.harvard.edu/abs/2017PhRvD..95j3012B}
}

@ARTICLE{Luo2016,
       author = {{Luo}, Jun and {Chen}, Li-Sheng and {Duan}, Hui-Zong and {Gong}, Yun-Gui and {Hu}, Shoucun and {Ji}, Jianghui and {Liu}, Qi and {Mei}, Jianwei and {Milyukov}, Vadim and {Sazhin}, Mikhail and {Shao}, Cheng-Gang and {Toth}, Viktor T. and {Tu}, Hai-Bo and {Wang}, Yamin and {Wang}, Yan and {Yeh}, Hsien-Chi and {Zhan}, Ming-Sheng and {Zhang}, Yonghe and {Zharov}, Vladimir and {Zhou}, Ze-Bing},
        title = "{TianQin: a space-borne gravitational wave detector}",
      journal = {Classical and Quantum Gravity},
         year = 2016,
        month = feb,
       volume = {33},
       number = {3},
          eid = {035010},
        pages = {035010},
          doi = {10.1088/0264-9381/33/3/035010},
archivePrefix = {arXiv},
       eprint = {1512.02076},
 primaryClass = {astro-ph.IM},
       adsurl = {https://ui.adsabs.harvard.edu/abs/2016CQGra..33c5010L}
}

@article{Hu2017,
    author = {Hu, Wen-Rui and Wu, Yue-Liang},
    title = {The Taiji Program in Space for gravitational wave physics and the nature of gravity},
    journal = {National Science Review},
    volume = {4},
    number = {5},
    pages = {685-686},
    year = {2017},
    month = {10},
    issn = {2095-5138},
    doi = {10.1093/nsr/nwx116},
    url = {https://doi.org/10.1093/nsr/nwx116},
    eprint = {https://academic.oup.com/nsr/article-pdf/4/5/685/31566708/nwx116.pdf},
}

@ARTICLE{Cheng2025,
       author = {{Cheng}, Ya-Ze and {Cao}, Yan and {Tang}, Yong},
        title = "{Effects of black hole environments on extreme mass-ratio hyperbolic encounters}",
      journal = {\prd},
         year = 2025,
        month = apr,
       volume = {111},
       number = {8},
          eid = {083010},
        pages = {083010},
          doi = {10.1103/PhysRevD.111.083010},
archivePrefix = {arXiv},
       eprint = {2411.03095},
 primaryClass = {gr-qc},
       adsurl = {https://ui.adsabs.harvard.edu/abs/2025PhRvD.111h3010C}
}

@ARTICLE{Yuan2025,
       author = {{Yuan}, Hao-Yu and {Lei}, Wei-Hua},
        title = "{The Multiband Emission of the Two-component Gamma-Ray Burst Jet Influenced by Progenitor Winds within the Accretion Disk of Active Galactic Nuclei}",
      journal = {\apj},
         year = 2025,
        month = jul,
       volume = {987},
       number = {2},
          eid = {167},
        pages = {167},
          doi = {10.3847/1538-4357/addbe4},
archivePrefix = {arXiv},
       eprint = {2408.06593},
 primaryClass = {astro-ph.HE},
       adsurl = {https://ui.adsabs.harvard.edu/abs/2025ApJ...987..167Y}
}

@ARTICLE{He2025,
       author = {{He}, Lei and {Liu}, Zheng-Yan and {Niu}, Rui and {Zhou}, Ming-Shen and {Zou}, Pu-Run and {Gao}, Bing-Zhou and {Liang}, Run-Duo and {Zhu}, Liang-Gui and {Wang}, Jian-Min and {Jiang}, Ning and {Cai}, Zhen-Yi and {Jiang}, Ji-an and {Dai}, Zi-Gao and {Yuan}, Ye-Fei and {Chen}, Yong-Jie and {Zhao}, Wen},
        title = "{A Systematic Search for AGN Flares in ZTF Data Release 23}",
      journal = {arXiv e-prints},
         year = 2025,
        month = jul,
          eid = {arXiv:2507.20232},
        pages = {arXiv:2507.20232},
          doi = {10.48550/arXiv.2507.20232},
archivePrefix = {arXiv},
       eprint = {2507.20232},
 primaryClass = {astro-ph.HE},
       adsurl = {https://ui.adsabs.harvard.edu/abs/2025arXiv250720232H}
}

@ARTICLE{Chen2025,
       author = {{Chen}, Ken and {Dai}, Zi-Gao},
        title = "{Observational Properties of Thermal Emission from Relativistic Jets Embedded in Active Galactic Nucleus Disks}",
      journal = {\apj},
         year = 2025,
        month = jul,
       volume = {987},
       number = {2},
          eid = {214},
        pages = {214},
          doi = {10.3847/1538-4357/addb48},
archivePrefix = {arXiv},
       eprint = {2505.16390},
 primaryClass = {astro-ph.HE},
       adsurl = {https://ui.adsabs.harvard.edu/abs/2025ApJ...987..214C}
}

@ARTICLE{Zhang2025,
       author = {{Zhang}, Shu-Rui and {Wang}, Yu and {Yuan}, Ye-Fei and {Tagawa}, Hiromichi and {Wei}, Yun-Feng and {Li}, Liang and {Cai}, Rong-Gen},
        title = "{S241125n: Binary Black Hole Merger Produces Short GRB in AGN Disk}",
      journal = {arXiv e-prints},
         year = 2025,
        month = may,
          eid = {arXiv:2505.10395},
        pages = {arXiv:2505.10395},
          doi = {10.48550/arXiv.2505.10395},
archivePrefix = {arXiv},
       eprint = {2505.10395},
 primaryClass = {astro-ph.HE},
       adsurl = {https://ui.adsabs.harvard.edu/abs/2025arXiv250510395Z}
}

@ARTICLE{Zhang2024,
       author = {{Zhang}, Hao-Hui and {Zhu}, Jin-Ping and {Yu}, Yun-Wei},
        title = "{Propagation of Gamma-Ray Burst Relativistic Jets in Active Galactic Nucleus Disks and Its Implication for Gamma-Ray Burst Detection}",
      journal = {\apj},
         year = 2024,
        month = nov,
       volume = {976},
       number = {1},
          eid = {63},
        pages = {63},
          doi = {10.3847/1538-4357/ad8139},
archivePrefix = {arXiv},
       eprint = {2406.10904},
 primaryClass = {astro-ph.HE},
       adsurl = {https://ui.adsabs.harvard.edu/abs/2024ApJ...976...63Z}
}

@ARTICLE{Kang2025,
       author = {{Kang}, Hoyoung D. and {Perna}, Rosalba and {Lazzati}, Davide and {Wang}, Yi-Han},
        title = "{The Cosmological Population of Gamma-Ray Bursts from the Disks of Active Galactic Nuclei}",
      journal = {The Open Journal of Astrophysics},
         year = 2025,
        month = mar,
       volume = {8},
          eid = {23},
        pages = {23},
          doi = {10.33232/001c.131902},
archivePrefix = {arXiv},
       eprint = {2412.17714},
 primaryClass = {astro-ph.HE},
       adsurl = {https://ui.adsabs.harvard.edu/abs/2025OJAp....8E..23K}
}

@ARTICLE{Fabj2025,
       author = {{Fabj}, Gaia and {Dittmann}, Alexander J. and {Cantiello}, Matteo and {Perna}, Rosalba and {Samsing}, Johan},
        title = "{Mapping the Outcomes of Stellar Evolution in the Disks of Active Galactic Nuclei}",
      journal = {\apj},
         year = 2025,
        month = mar,
       volume = {981},
       number = {1},
          eid = {16},
        pages = {16},
          doi = {10.3847/1538-4357/ada896},
archivePrefix = {arXiv},
       eprint = {2408.16050},
 primaryClass = {astro-ph.GA},
       adsurl = {https://ui.adsabs.harvard.edu/abs/2025ApJ...981...16F}
}

@ARTICLE{LIGO2025,
       author = {{The LIGO Scientific Collaboration} and {the Virgo Collaboration} and {the KAGRA Collaboration} and {Abac}, A.~G. and {Abouelfettouh}, I. and {Acernese}, F. and {Ackley}, K. and {Adamcewicz}, C. and {Adhicary}, S. and {Adhikari}, D. and {Adhikari}, N. and {Adhikari}, R.~X. and {Adkins}, V.~K. and {Afroz}, S. and {Agapito}, A. and {Agarwal}, D. and {Agathos}, M. and {Aggarwal}, N. and {Aggarwal}, S. and {Aguiar}, O.~D. and {Ahrend}, I. -L. and {Aiello}, L. and {Ain}, A. and {Ajith}, P. and {Akutsu}, T. and {Albanesi}, S. and {Ali}, W. and {Al-Kershi}, S. and {All{\'e}n{\'e}}, C. and {Allocca}, A. and {Al-Shammari}, S. and {Altin}, P.~A. and {Alvarez-Lopez}, S. and {Amar}, W. and {Amarasinghe}, O. and {Amato}, A. and {Amicucci}, F. and {Amra}, C. and {Ananyeva}, A. and {Anderson}, S.~B. and {Anderson}, W.~G. and {Andia}, M. and {Ando}, M. and {Andr{\'e}s-Carcasona}, M. and {Andri{\'c}}, T. and {Anglin}, J. and {Ansoldi}, S. and {Antelis}, J.~M. and {Antier}, S. and {Aoumi}, M. and {Appavuravther}, E.~Z. and {Appert}, S. and {Apple}, S.~K. and {Arai}, K. and {Araujo Alvarez}, C. and {Araya}, A. and {Araya}, M.~C. and {Arca Sedda}, M. and {Areeda}, J.~S. and {Aritomi}, N. and {Armato}, F. and {Armstrong}, S. and {Arnaud}, N. and {Arogeti}, M. and {Aronson}, S.~M. and {Arun}, K.~G. and {Ashton}, G. and {Aso}, Y. and {Asprea}, L. and {Assiduo}, M. and {Assis de Souza Melo}, S. and {Aston}, S.~M. and {Astone}, P. and {Attadio}, F. and {Aubin}, F. and {AultONeal}, K. and {Avallone}, G. and {Avila}, E.~A. and {Babak}, S. and {Badger}, C. and {Bae}, S. and {Bagnasco}, S. and {Baiotti}, L. and {Bajpai}, R. and {Baka}, T. and {Baker}, A.~M. and {Baker}, K.~A. and {Baker}, T. and {Baldi}, G. and {Baldicchi}, N. and {Ball}, M. and {Ballardin}, G. and {Ballmer}, S.~W. and {Banagiri}, S. and {Banerjee}, B. and {Bankar}, D. and {Baptiste}, T.~M. and {Baral}, P. and {Baratti}, M. and {Barayoga}, J.~C. and {Barish}, B.~C. and {Barker}, D. and {Barman}, N. and {Barneo}, P. and {Barone}, F. and {Barr}, B. and {Barsotti}, L. and {Barsuglia}, M. and {Barta}, D. and {Bartoletti}, A.~M. and {Barton}, M.~A. and {Bartos}, I. and {Basalaev}, A. and {Bassiri}, R. and {Basti}, A. and {Bawaj}, M. and {Baxi}, P. and {Bayley}, J.~C. and {Baylor}, A.~C. and {Baynard}, II, P.~A. and {Bazzan}, M. and {Bedakihale}, V.~M. and {Beirnaert}, F. and {Bejger}, M. and {Belardinelli}, D. and {Bell}, A.~S. and {Bellie}, D.~S. and {Bellizzi}, L. and {Benoit}, W. and {Bentara}, I. and {Bentley}, J.~D. and {Ben Yaala}, M. and {Bera}, S. and {Bergamin}, F. and {Berger}, B.~K. and {Bernuzzi}, S. and {Beroiz}, M. and {Berry}, C.~P.~L. and {Bersanetti}, D. and {Bertheas}, T. and {Bertolini}, A. and {Betzwieser}, J. and {Beveridge}, D. and {Bevilacqua}, G. and {Bevins}, N. and {Bhandare}, R. and {Bhatt}, R. and {Bhattacharjee}, D. and {Bhattacharyya}, S. and {Bhaumik}, S. and {Bhagwat}, S. and {Biancalana}, V. and {Bianchi}, A. and {Bilenko}, I.~A. and {Billingsley}, G. and {Binetti}, A. and {Bini}, S. and {Binu}, C. and {Biot}, S. and {Birnholtz}, O. and {Biscoveanu}, S. and {Bisht}, A. and {Bitossi}, M. and {Bizouard}, M. -A. and {Blaber}, S. and {Blackburn}, J.~K. and {Blagg}, L.~A. and {Blair}, C.~D. and {Blair}, D.~G. and {Bode}, N. and {Boettner}, N. and {Boileau}, G. and {Boldrini}, M. and {Bolingbroke}, G.~N. and {Bolliand}, A. and {Bonavena}, L.~D. and {Bondarescu}, R. and {Bondu}, F. and {Bonilla}, E. and {Bonilla}, M.~S. and {Bonino}, A. and {Bonnand}, R. and {Borchers}, A. and {Borhanian}, S. and {Boschi}, V. and {Bose}, S. and {Bossilkov}, V. and {Bothra}, Y. and {Boudon}, A. and {Bourg}, L. and {Bouyer}, G. and {Boyle}, M. and {Bozzi}, A. and {Bradaschia}, C. and {Brady}, P.~R. and {Branch}, A. and {Branchesi}, M. and {Braun}, I. and {Briant}, T. and {Brillet}, A.},
        title = "{GW231123: a Binary Black Hole Merger with Total Mass 190-265 $M_{\odot}$}",
      journal = {arXiv e-prints},
         year = 2025,
        month = jul,
          eid = {arXiv:2507.08219},
        pages = {arXiv:2507.08219},
          doi = {10.48550/arXiv.2507.08219},
archivePrefix = {arXiv},
       eprint = {2507.08219},
 primaryClass = {astro-ph.HE},
       adsurl = {https://ui.adsabs.harvard.edu/abs/2025arXiv250708219T}
}

@BOOK{Meier2012,
       author = {{Meier}, David L.},
        title = "{Black Hole Astrophysics: The Engine Paradigm}",
         year = 2012,
          doi = {10.1007/978-3-642-01936-4},
       adsurl = {https://ui.adsabs.harvard.edu/abs/2012bhae.book.....M}
}

@ARTICLE{Cao2022,
       author = {{Cao}, Xinwu and {Gu}, Wei-Min},
        title = "{A Supercritical Accretion Disk with Radiation-driven Outflows}",
      journal = {\apj},
         year = 2022,
        month = sep,
       volume = {936},
       number = {2},
          eid = {141},
        pages = {141},
          doi = {10.3847/1538-4357/ac8980},
archivePrefix = {arXiv},
       eprint = {2208.06088},
 primaryClass = {astro-ph.HE},
       adsurl = {https://ui.adsabs.harvard.edu/abs/2022ApJ...936..141C}
}

@ARTICLE{Bellovary2016,
       author = {{Bellovary}, Jillian M. and {Mac Low}, Mordecai-Mark and {McKernan}, Barry and {Ford}, K.~E. Saavik},
        title = "{Migration Traps in Disks around Supermassive Black Holes}",
      journal = {\apjl},
         year = 2016,
        month = mar,
       volume = {819},
       number = {2},
          eid = {L17},
        pages = {L17},
          doi = {10.3847/2041-8205/819/2/L17},
archivePrefix = {arXiv},
       eprint = {1511.00005},
 primaryClass = {astro-ph.GA},
       adsurl = {https://ui.adsabs.harvard.edu/abs/2016ApJ...819L..17B}
}

@ARTICLE{Edgar2004,
       author = {{Edgar}, Richard},
        title = "{A review of Bondi-Hoyle-Lyttleton accretion}",
      journal = {\nar},
         year = 2004,
        month = sep,
       volume = {48},
       number = {10},
        pages = {843-859},
          doi = {10.1016/j.newar.2004.06.001},
archivePrefix = {arXiv},
       eprint = {astro-ph/0406166},
 primaryClass = {astro-ph},
       adsurl = {https://ui.adsabs.harvard.edu/abs/2004NewAR..48..843E}
}

@ARTICLE{Liu2024,
       author = {{Liu}, Jun-Rong and {Wang}, Yi-Lin and {Wang}, Jian-Min},
        title = "{Accretion-modified Stars in Accretion Disks of Active Galactic Nuclei: Observational Characteristics in Different Regions of the Disks}",
      journal = {\apj},
         year = 2024,
        month = jul,
       volume = {969},
       number = {1},
          eid = {37},
        pages = {37},
          doi = {10.3847/1538-4357/ad463a},
archivePrefix = {arXiv},
       eprint = {2405.02855},
 primaryClass = {astro-ph.HE},
       adsurl = {https://ui.adsabs.harvard.edu/abs/2024ApJ...969...37L}
}

@ARTICLE{Chen2023YiXian,
       author = {{Chen}, Yi-Xian and {Jiang}, Yan-Fei and {Goodman}, Jeremy and {Ostriker}, Eve C.},
        title = "{3D Radiation Hydrodynamic Simulations of Gravitational Instability in AGN Accretion Disks: Effects of Radiation Pressure}",
      journal = {\apj},
         year = 2023,
        month = may,
       volume = {948},
       number = {2},
          eid = {120},
        pages = {120},
          doi = {10.3847/1538-4357/acc023},
archivePrefix = {arXiv},
       eprint = {2302.10868},
 primaryClass = {astro-ph.HE},
       adsurl = {https://ui.adsabs.harvard.edu/abs/2023ApJ...948..120C}
}

@ARTICLE{Balbus1991,
       author = {{Balbus}, Steven A. and {Hawley}, John F.},
        title = "{A Powerful Local Shear Instability in Weakly Magnetized Disks. I. Linear Analysis}",
      journal = {\apj},
         year = 1991,
        month = jul,
       volume = {376},
        pages = {214},
          doi = {10.1086/170270},
       adsurl = {https://ui.adsabs.harvard.edu/abs/1991ApJ...376..214B}
}

@ARTICLE{Lynden-Bell1974,
       author = {{Lynden-Bell}, D. and {Pringle}, J.~E.},
        title = "{The evolution of viscous discs and the origin of the nebular variables.}",
      journal = {\mnras},
         year = 1974,
        month = sep,
       volume = {168},
        pages = {603-637},
          doi = {10.1093/mnras/168.3.603},
       adsurl = {https://ui.adsabs.harvard.edu/abs/1974MNRAS.168..603L}
}

@ARTICLE{Lynden-Bell1969,
       author = {{Lynden-Bell}, D.},
        title = "{Galactic Nuclei as Collapsed Old Quasars}",
      journal = {\nat},
         year = 1969,
        month = aug,
       volume = {223},
       number = {5207},
        pages = {690-694},
          doi = {10.1038/223690a0},
       adsurl = {https://ui.adsabs.harvard.edu/abs/1969Natur.223..690L}
}

@ARTICLE{Shakura1973,
       author = {{Shakura}, N.~I. and {Sunyaev}, R.~A.},
        title = "{Black holes in binary systems. Observational appearance.}",
      journal = {\aap},
         year = 1973,
        month = jan,
       volume = {24},
        pages = {337-355},
       adsurl = {https://ui.adsabs.harvard.edu/abs/1973A&A....24..337S}
}

@ARTICLE{Zeldovich1964,
       author = {{Zeldovich}, Ya. B.},
        title = "{The Fate of a Star and the Evolution of Gravitational Energy Upon Accretion}",
      journal = {Soviet Physics Doklady},
         year = 1964,
        month = sep,
       volume = {9},
        pages = {195},
       adsurl = {https://ui.adsabs.harvard.edu/abs/1964SPhD....9..195Z}
}
\bibliographystyle{aasjournal}
\end{document}